\documentclass[review]{elsarticle}

\usepackage{lineno,hyperref}
\usepackage{amsmath}
\usepackage{amssymb}
\usepackage{bm}
\usepackage{subcaption}
\usepackage{xcolor}
\modulolinenumbers[5]

\journal{arXiv}

\def\bu{\boldsymbol{u}}
\def\bs{\boldsymbol{s}}
\def\bv{\boldsymbol{v}}
\def\bf{\boldsymbol{f}}

\def\bpsi{\boldsymbol{\Psi}}
\def\bphi{\boldsymbol{\phi}}

\def\bphi{\boldsymbol{\Phi}}

\begin{document}

\begin{frontmatter}

\title{Sensitivity analysis of chaotic systems using a frequency-domain shadowing approach}

\author{Kyriakos D. Kantarakias}
\author{George Papadakis}
\address{Department of Aeronautics, Imperial College London, SW7 2AZ, U.K.}

\begin{abstract}
We present a frequency-domain method for computing the sensitivities of time-averaged quantities of chaotic systems with respect to input parameters. Such sensitivities cannot be computed by conventional adjoint analysis tools, because the presence of positive Lyapunov exponents leads to exponential growth of the adjoint variables. The proposed method is based on the least-square shadowing (LSS) approach \cite{WANG2014210}, that formulates the evaluation of sensitivities as an optimisation problem, thereby avoiding the exponential growth of the solution. However, all existing formulations of LSS (and its variants) are in the time domain and the computational cost scales with the number of positive Lyapunov exponents. In the present paper, we reformulate the LSS method in the Fourier space using harmonic balancing.  The new method is tested on the Kuramoto-Sivashinski system and the results match with those obtained using the standard time-domain formulation. Although the cost of the direct solution is independent of the number of positive Lyapunov exponents, storage and computing requirements grow rapidly with the size of the system. To mitigate these requirements, we propose a resolvent-based iterative approach that needs much less storage. Application to the Kuramoto-Sivashinski system gave accurate results with very low computational cost. The method is applicable to large systems and paves the way for application of the resolvent-based shadowing approach to turbulent flows. Further work is needed to assess its performance and scalability. 
\end{abstract}


\end{frontmatter}

\linenumbers

\section{Introduction}

Optimisation of engineering devices is based of the definition of an objective function, usually a time-average quantity $\overline{J}(\bs)$, and the evaluation of the problem parameters, $\bs$, that minimise or maximise this function, depending on the application. 
During the optimisation process, the gradient of the objective function with respect to the parameters $d\overline{J}(\bs)/d \bs$ (also known as sensitivity) is usually required. This is obtained by solving the tangent equation (or the adjoint equations for multiple parameters). In either case, the equations are obtained by linearising the governing non-linear set describing the system around the solution obtained for the reference values of the parameters, $\bs$. The tangent (or adjoint) equations are then integrated forward (or backward) in time respectively to obtain the desired sensitivities. 

The aforementioned approach works very well when the governing set of equations describing the system is steady, in which case the solution is a point in phase space. When the evolution is however unsteady, and in particular when the system exhibits chaotic behaviour, this process fails. The reason is that chaotic systems have one or more positive Lyapunov exponents (PLEs), thus two solution trajectories starting from the same initial conditions and evaluated at $\bs$ and $\bs+\delta \bs$ deviate from each other, leading to exponentially growing sensitivities, as explained in \cite{Leapaper}. For this reason for example, model predictive control algorithms for transitional or turbulent flows employ the receding horizon approach, whereby the optimisation is performed over a receding window of finite time \cite{bewley_moin_temam_2001,xiao_papadakis_2019}, thus sensitivities remain bounded and reliable. 

Several approaches that can compute useful sensitivities in chaotic systems have been proposed. These are based on ensemble schemes \cite{Eyink_2004}, the fluctuation dissipation theorem \cite{kubo_1966}, the Fokker-Planck equation \cite{Thuburn2005}, cumulant expansions \cite{CRASKE2019243}, or unsteady periodic orbits \cite{Lasagna_2018}. One of the most promising approaches is Least Squares Shadowing (LSS) \cite{WANG2014210,Wang2014ConvergenceAverages,blonigan_airfoil}, which is based on the shadowing lemma  \cite{Pilyugin_1999,Bowen1975-LimitDiffeomorphisms}. For uniformly hyperbolic systems, this lemma guarantees the existence of a solution trajectory evaluated at $\bs+\delta \bs$ that shadows, i.e.\ remains close to, the reference trajectory evaluated at  $\bs$. This regularises the problem, avoids the exponential growth, and results in meaningful sensitivities. 

This lemma is also central in establishing trust into the statistics of  numerical solutions of chaotic systems. Due to round off errors, a computed trajectory will deviate from the true trajectory of the system (starting from the same initial condition). The shadowing lemma guarantees the existence of a true trajectory (with different initial condition) that will shadow the numerical one  \cite{Hammel_et_al_1987,Sauer_Yorke_1991,Sauer_et_al_1997}, thus the statistics of the computed solution can be trusted. Recent work however \cite{CHANDRAMOORTHY2021110389} has shown that the shadowing trajectories may not be physical, thus casting doubt on this central premise. Several chaotic  systems (one dimensional perturbed tent maps) were examined; the shadowing trajectories were found to be physical in one system and non-physical in others. This finding raises important fundamental questions, such as under what conditions shadowing solutions are physical, what happens for higher dimensional systems etc. In this paper we take the standard view, that shadowing solutions are physical, and numerical simulations of chaotic systems reproduce the true statistics.  

Variants of the original LSS method include the  Multiple Shooting Shadowing (MSS) \cite{bloniganMSS,Shawki_Papadakis_2019,Kantarakias_Shawki_Papadakis_2020} and the non-intrusive Least Squares Shadowing (NILSS), \cite{NI201756,NI2019690}. Both methods can be applied to large systems, but their computational cost scales with the number of PLEs. This limitation restricts the application to systems with relatively small or moderate number of PLE's. For example, NILSS has been applied successfully to 2D flow (backward facing step with 14 PLEs \cite{NI201756}) and to 3D flows (minimal channel flow unit at $Re_\tau=140$ with 150-160 PLEs \cite{BLONIGAN2017803}, and flow around a cylinder at Re=525 with less than 30 PLEs \cite{ni_2019}). 

The number of PLEs and how it changes with the system parameters is therefore of critical importance for the application of the method. In the area of turbulent flows, the Lyapunov spectrum has been computed for low-Reynolds channel  \cite{keefe_moin_kim_1992} and weakly chaotic Taylor-Couette flows \cite{vastano_moser_1991}. A more recent study \cite{Hassanaly_Raman_2019} has investigated the variation of the spectrum with Reynolds number for forced homogeneous isotropic turbulence (HIT). The Reynolds numbers examined (based on the Taylor microscale) were $Re_\lambda=15.5, 21.3, 25.6$ (note that these values are considered very small for engineering applications). The number of PLEs were found to be about 25, 60 and 100 respectively (see figure 2 of \cite{Hassanaly_Raman_2019}). A fourth value of $Re_\lambda=37.7$ was also considered, but it was not possible to find the number of PLEs because of the slow decay rate of the spectrum. The latter was found to follow a power-law, $\lambda_i \approx \alpha(i-1)^\beta+\lambda_1$, with the exponent $\beta$ in the region 0.81-0.85 (with the smaller value for the higher $Re_\tau$).  The maximum LE, $\lambda_1$, is expected to scale with the inverse of the Kolmogorov time scale, $\tau_{\eta}$, see theoretical arguments in \cite{Crisanti_et_al_1993}. This was tested in \cite{Mohan_et_al_2017} for HIT and it was found that $\lambda_1 \tau_{\eta}$ is not constant, but instead grows with $Re_\lambda$ following a power-law,  $\lambda_1 \tau_{\eta} \sim Re_\lambda^\gamma$.

The above scaling of $\lambda_1$ and the fact that the decay rate of the Lyapunov spectrum decreases with Reynolds number means that the number of PLEs grows very rapidly as Reynolds increases. Thus, alternative approaches are required to make LSS and its variants applicable to complex flows of engineering interest. One such approach relies on the understanding of the underlying physical processes. For example, it is well known that momentum transfer is dominated by large scale structures, thus smaller scales (that are responsible for the largest LEs) can be filtered out and their effect modelled, hopefully without loss of accuracy in sensitivity. This is exactly what large eddy simulations (LES) are designed for \cite{pope_2000}. In standard LES, the equations are filtered in space, but temporal filtering (with filter time scale $\Delta$) is also possible \cite{Pruett_2008}. In the limit of $\Delta \to \infty$, the temporally-averaged LES (TLES) equations tend to the standard Reynolds-Averaged Navier-Stokes (RANS) equations (section 2.3 of \cite{Pruett_2008}). We conjecture therefore that as $\Delta$ increases, application of LSS to the TLES equations will recover the sensitivites predicted by the tangent (or adjoint) method applied to RANS. Thus, parameter $\Delta$ bridges two limits, LSS applied to unfiltered Navier-Stokes ($\Delta=0$) and the RANS equations ($\Delta \to \infty$). As $\Delta$ increases, the number of PLEs decreases and the problem becomes better conditioned. However, accuracy is traded for computational efficiency, because as $\Delta$ increases the effect of more scales needs to be modelled.

Although the above approach mitigates the rapid growth of the computational cost of LSS (and also provides a useful conceptual framework that bridges two limits), it relies on accurate modelling of the filtered scales. In the present paper, we follow a different approach. All existing formulations of LSS and its variants have been derived in the time domain. If however the LSS is formulated in the frequency domain, the exponential separation of the trajectories for $\bs$ and $\bs+\delta \bs$ does not appear explicitly. Of course time- and frequency-domain formulations are equivalent, but as will be seen, the latter formulation allows us to gain deep physical insight and also is amenable to iterative solution algorithms that are not possible with the former. Frequency domain approaches have been applied from the perspective of linear \cite{mckeon_sharma_2010} and non-linear input-output analysis \cite{rigas_sipp_colonius_2021}, the frequency response of periodically time-varying base flows \cite{padovan_otto_rowley_2020}, or model-based design of transverse wall oscillations for turbulent drag reduction in a channel flow \citep{moarref_jovanovic_2012}.  Similarities and differences with existing frequency-domain approaches are discussed throughout the manuscript.

The paper is organised as follows. Section \ref{Shadowing Basics} sets scene and presents the standard LSS algorithm in the time domain. The formulation of the algorithm in the frequency domain is derived in section \ref{sensitivity analysis} followed by application to the Kuramoto-Sivasinsky equation in section \ref{sec:app_Kuramoto_Sivashinsky}. A resolvent-based iterative algorithm to solve the resulting system is presented in section  \ref{Simplified Fourier} and the results are further analysed in section \ref{Resolvent Decoupled}. We conclude in section \ref{conclusions chapter}. 

\section{Sensitivity analysis of chaotic systems using the shadowing approach} \label{Shadowing Basics}
Consider a dynamical system governed by a set of ordinary differential equations of the form,
\begin{equation}
\label{ODE_original}
\begin{aligned}
&\frac{d \bu}{dt} = f(\bu;\bs)\\
&\bu(0; \bs) = \bu_0(\bs),
\end{aligned}
\end{equation}
\noindent where $\bu(t;\bs) \in \mathbb{R}^{N_{\bu}}$ is the vector of state variables and $\bs \in \mathbb{R}^{N_{\bs}}$  is the set of control parameters that define the dynamics of the system.  System \eqref{ODE_original} can arise for example after spatial discretisation of a set of conservation laws that describe mathematically the problem under investigation. We assume that the vector field $ \bf(\bu; \bs): \mathbb{R}^{N_{\bu}} \times  \mathbb{R}^{N_{\bs}} \to \mathbb{R}^{N_{\bu}}$ varies smoothly with $\bu$ and $\bs$. 

In many applications we are interested in evaluating the sensitivity of a time-averaged quantity $\overline{J}(\bs): \mathbb{R}^{N_{\bs}} \to \mathbb{R}$,  \begin{equation}
\label{QoI}
    \overline{J}(\bs) = \lim_{T \to \infty} \frac{1}{T} \int_0^T J(\bu;\bs) dt,
\end{equation}
\noindent to the parameters $\bs$. For example, in the area of aerodynamics, $\overline{J}(\bs)$ can be the drag coefficient and $\bs$ the set of variables  that describe the shape of an airfoil. The gradient of $\overline{J}(\bs)$ with respect to $\bs$ is defined as 
\begin{equation}
\begin{aligned}
\frac{d \overline{J}}{d \bs}  & =  \lim_{ \delta \bs \to 0} \frac{\overline{J}(\bs+ \delta \bs)-\overline{J}(\bs)} {\delta \bs} \\
& =\lim_{ \delta \bs \to 0} \lim_{T \to \infty} \frac{1}{T} \int_0^T \frac{J \left(\bu(t; \bs+ \delta \bs);\bs+ \delta \bs \right) - J \left(\bu(t; \bs); \bs \right)}{ \delta \bs}dt
\end{aligned}
\label{eq:tangent_sensitivity}
\end{equation}
In chaotic systems, the limit and differentiation operations do not commute, i.e.\
\begin{equation}
\begin{aligned}
& \frac{d \overline{J}}{d \bs}  \ne
\lim_{T \to \infty}  \lim_{ \delta \bs \to 0} 
\frac{1}{T} \int_0^T \frac{J \left(\bu(t; \bs+ \delta \bs);\bs+ \delta \bs \right) - J \left(\bu(t; \bs); \bs \right)}{ \delta \bs}dt \\
& \left [ =\lim_{T \to \infty}  \frac{1}{T} \int_0^T \frac{dJ \left(\bu(t; \bs); \bs\right)}{ d \bs}dt =
\lim_{T \to \infty}  \frac{1}{T} \int_0^T \left (\frac{\partial J}{\partial \bu} \bv+ \frac{\partial J}{\partial \bs} \right ) dt
\right ],
\end{aligned}
\label{eq:non_commutation}
\end{equation}
where 
\begin{equation}
   \bv(t) = \frac{d\bu}{ d \bs} = \lim_{ \delta \bs \to 0} \frac{\bu(t; \bs+ \delta \bs) - \bu(t; \bs)}{ \delta \bs} 
   \label{eq:definition-v-sensitivity}
\end{equation}
\noindent is the sensitivity of the solution $\bu(t; \bs)$ to a change $\delta \bs$ of $\bs$. The reason is that chaotic systems have one or more PLEs. This means that the distance in phase space between  $\bu(t; \bs+ \delta \bs)$ and $\bu(t; \bs)$, i.e. the Euclidean norm of the vector $\bu(t; \bs+ \delta \bs) - \bu(t; \bs)$ that appears in the nominator of \eqref{eq:definition-v-sensitivity},  grows exponentially at rate $\sim e^{\lambda_1 t}$, where $\lambda_1$ is the maximum of these exponents \cite{Leapaper,Thuburn2005,WANG20131}. Thus, the quantity $ \frac{1}{T} \int_0^T \left (\frac{\partial J}{\partial \bu} \bv+ \frac{\partial J}{\partial \bs} \right )$ that appears on the right hand side of \eqref{eq:non_commutation} diverges as  $T\to \infty$. On the other hand, assuming that $\overline{J}(\bs)$ varies smoothly with $\bs$, the sensitivity $\frac{d \overline{J}}{d \bs}$ is finite.

If the dynamical system \eqref{ODE_original} is uniformly hyperbolic, the shadowing lemma \cite{Pilyugin1999ShadowingSystems} guarantees the existence of a solution trajectory evaluated at $\bs+ \delta \bs$ that remains always close, i.e. shadows indefinitely, the reference trajectory $\bu( t; \bs)$. We denote this shadowing trajectory as $\bu( \tau(t); \bs+ \delta \bs)$, where $\tau(t)$ is an appropriate time transformation. The LSS  method, proposed in \cite{WANG2014210}, computes the shadowing trajectory by minimising the distance between $\bu(\tau(t); \bs+ \delta \bs)$ and $\bu(t; \bs)$ in a least squares sense, i.e.\ 
\begin{subequations}
\begin{align}
& \min_{\bu,\tau} \frac{1}{2} \int_0^T  \left( \left\lVert \bu \left( \tau(t); \bs+ \delta \bs \right) - \bu(t;\bs) \right\rVert^2 + \alpha^2 \left\lVert \frac{d\tau}{dt} -1 \right\rVert ^2 \right) dt \mbox{  }  s.t. \label{eq:cost_function_1} \\
&\frac{d \bu}{dt} = \bf \left( \bu \left (\tau(t); \bs+ \delta \bs \right) \right),
\end{align}
\label{eq:minimisation_1}
\end{subequations}
\noindent where $\alpha^2$ is a constant parameter. 
Taking the limit $\delta \bs \to 0$, leads to the following linear minimisation problem,
\begin{subequations}
\begin{align}
& \min_{\bv,\eta} \frac{1}{2} \int_0^T \left( \| \bv(t) \|^2 + \alpha^2 \|\eta \|^2 \right) dt \mbox{  }  s.t. \label{eq:cost_function_2}  \\
&\frac{d \bv}{dt} = \frac{\partial \bf}{\partial \bu} \bv + \frac{\partial \bf}{\partial \bs} + \eta(t) \bf,
\end{align}
\label{eq:minimisation_2}
\end{subequations}
\noindent where
\begin{subequations}
\begin{align}
& \bv(t) = \frac{d}{d \bs} \bu(\tau(t);\bs) \\
&\eta(t) = \frac{d}{d \bs} \left( \frac{d \tau}{dt} \right).
\end{align}
\label{v and eta}
\end{subequations}
\noindent The second term within the cost functions   \eqref{eq:cost_function_1} and \eqref{eq:cost_function_2} penalises the deviation of $\tau(t)$ from $t$. A high value of $\alpha^2$ results in a small deviation (heavy penalisation), while a small value to light penalisation. The solution of \eqref{eq:minimisation_2} for $\alpha^2=0$, leads to the orthogonality condition between the vectors $\bf(\bu;s)$ and $\bv(\bu;s)$ at each point along the trajectory, i.e.\ $ \langle  \bf(\bu;\bs) \bv(\bu;\bs) \rangle  = 0$, a constraint from which $\eta(t)$ can be obtained \cite{bloniganMSS}. Thus, problem \eqref{eq:minimisation_2} becomes
\begin{subequations}
\begin{align}
& \min_{\bv,\eta}\frac{1}{2} \int_0^T  \| \bv(t) \|^2 dt  \mbox{  } s.t.  \label{eq:cost_function_3} \\
&\frac{d \bv}{dt} = \frac{\partial \bf}{\partial \bu} \bv + \frac{\partial \bf}{\partial \bs} + \eta(t) \bf \label{eq:ODE_tangent_dilation}\\
& \langle  \bf(\bu;\bs) \bv(\bu;\bs) \rangle  = 0  \label{eq:orthogonality}
\end{align}
\label{eq:minimisation_3}
\end{subequations}
From the solution $\bv_{lss}(t)$ and $\eta_{lss}(t)$ of \eqref{eq:minimisation_3}, the sensitivity $\frac{d \overline{J}}{d \bs}$ can be easily computed \cite{WANG2014210}.

\section{Formulation of the shadowing algorithm in Fourier space} \label{sensitivity analysis}
As mentioned in the Introduction, all existing methods solve the minimisation problem \eqref{eq:minimisation_3} in the time domain. In this section, we formulate the problem in the frequency domain, i.e.\ in Fourier space, and seek a solution that remains bounded. 

To this end, we consider a reference trajectory $\bu(t; \bs)$ of length $T$ and assume that the solution of the minimisation problem \eqref{eq:minimisation_3} is periodic with period $T$. Thus, it can be written in terms of Fourier series as,
\begin{equation}
\label{Fourier Series 1}
\bv(t) = \sum_{ k = -\infty}^{+\infty} \hat{\bv}_{k} e^{i k\omega_0  t}, \mbox{   } \eta(t) = \sum_{ k = -\infty}^{+\infty} \hat{\eta}_{k}  e^{i k\omega_0 t},
\end{equation}
where $\hat{\bv}_{k},\hat{\eta}_{k}$ denote the Fourier coefficients, $\omega_0 = \frac{2 \pi}{T}$ is the fundamental angular frequency and the index $k$ characterises the harmonics with frequencies $\omega_k=k\omega_0$.  We assume similar series expansions for the Jacobian $\frac{\partial \bf}{\partial \bu}(t)$ and $\bf(t)$, 
\begin{equation}
\frac{\partial \bf}{\partial \bu}(t)= \sum_{ k = -\infty}^{+\infty}  \left( \widehat{\frac{\partial \bf}{\partial \bu}} \right)_k e^{i k\omega_0  t}, \mbox{   } \bf(t) = \sum_{ k = -\infty}^{+\infty} \hat{\bf}_{k}  e^{i k\omega_0 t}.
\label{Fourier Series 2}
\end{equation}
The matrix-vector product $\frac{\partial \bf}{\partial \bu}(t) \bv(t) $ can be written as
\begin{equation}
\frac{\partial \bf}{\partial \bu} \bv=
\sum_{ k = -\infty}^{+\infty}  \left( \widehat{\frac{\partial \bf}{\partial \bu} \bv} \right) _k e^{i k\omega_0  t},
\label{Fourier Series 3}
\end{equation}
where
\begin{equation}
\left( \widehat{\frac{\partial \bf}{\partial \bu} \bv}\right)_k=
\sum_{l = -\infty}^{+\infty} \left( \widehat{\frac{\partial \bf}{\partial \bu}}\right)_{k-l} \hat{\bv}_l,
\label{Fourier Series 4}
\end{equation}
which is the convolution sum between the Fourier coefficients of $\frac{\partial \bf}{\partial \bu}(t)$ and $\bv(t)$. Similarly, the left hand side of the orthogonality condition \eqref{eq:orthogonality} can be expanded as
\begin{equation}
\langle  \bf(\bu;\bs) \bv(\bu;\bs) \rangle  = \bf^\top \bv = \sum_{ k = -\infty}^{+\infty}  \left(\widehat{\bf^\top \bv}\right)_k e^{i k\omega_0  t}, 
\label{Fourier Series 5}
\end{equation}
where
\begin{equation}
\left(\widehat{\bf^\top \bv}\right)_k=\sum_{l = -\infty}^{+\infty} \hat{\bf}_{k-l}^\top \hat{\bv}_l,
\label{Fourier Series 6}
\end{equation}
and the notation $()^\top$ denotes the transpose operation. In the above two expressions, we have assumed that the weighting matrix associated with the inner product is the identity matrix, but the analysis below can be easily generalised to an inner product defined as $\langle  \bf(\bu;\bs) \bv(\bu;\bs) \rangle  = \bf^\top \mathcal{Q} \bv$. Finally,
\begin{equation}
\eta(t) \bf(t)=\sum_{ k = -\infty}^{+\infty}  \left(\widehat{\eta \bf}\right)_k e^{i k\omega_0  t} \quad \mbox{with} \quad  \left(\widehat{\eta \bf}\right)_k= \sum_{l = -\infty}^{+\infty} \hat{\bf}_{k-l} \hat{\eta}_{l} .
\label{Fourier Series 7}
\end{equation}

In practise, the range of the index $k$ is truncated to lie within the interval $[-q, q]$. For example, if the reference trajectory is sampled every $\Delta t$, $q=\frac{T}{2 \Delta t}-1$.  The frequency spectrum of $\bu(t; \bs)$ can also indicate the number of spectral coefficients that must be retained. A finite $q$ amounts to applying a sharp spectral cut-off filter to the above expansions, where all coefficients with $\lvert k \rvert, \lvert l \rvert$ or $\lvert k-l \rvert >q$ are set equal to $0$. 

Introducing the finite spectral representations to \eqref{eq:ODE_tangent_dilation} and \eqref{eq:orthogonality}, yields the following block set of equations for the $k$-th pair of coefficients $ \hat{\bv}_{k}, \hat{\eta}_{k}$,
\begin{equation}
i k\omega_0 \mathcal{I}_u
\begin{bmatrix}
  \hat{\bv}_{k} \\
 \hat{\eta}_{k}
\end{bmatrix}  
- \sum_{l= -q}^q T_{k-l}  
\begin{bmatrix}
  \hat{\bv}_{l} \\
 \hat{\eta}_{l}
\end{bmatrix}  
= 
\begin{bmatrix}
  \frac{d\hat{\bf}_k}{d \bs} \\
 0
\end{bmatrix}  
\label{eq:k-th_equation}
\end{equation} 
where
\begin{gather}
 \mathcal{I}_{u}
 =
  \begin{bmatrix}
   \mathcal{I}_{N_u} &   0 \\
   0 &   0 
   \end{bmatrix},
\end{gather}
\begin{gather}
T_{m}=
  \begin{bmatrix}
  \left( \widehat{\frac{\partial \bf}{\partial \bu}}\right)_{m}  &
  \hat{\bf}_m \\
   \hat{\bf}_{m}^\top  &
   0 
   \end{bmatrix},
   \label{eq:matrix_T_m}
\end{gather}
and $ \mathcal{I}_{N_u}$ is the identity matrix of dimension $N_u$. Each block consists of $N_u+1$ equations and there in total $2q+1$ blocks, resulting in $(N_u+1)\times(2q+1)$ equations and unknowns. 

Due to the sharp spectral cut-off  filter mentioned earlier, the starting and final values of index $l$ in the summation is slightly modified when $k\ne 0$. For example, for $k=-q$ equation \eqref{eq:k-th_equation} becomes
\begin{equation}
i (-q)\omega_0 \mathcal{I}_u
\begin{bmatrix}
  \hat{\bv}_{-q} \\
 \hat{\eta}_{-q}
\end{bmatrix}  
- \sum_{l= -q}^0 T_{-q-l}  
\begin{bmatrix}
  \hat{\bv}_{l} \\
 \hat{\eta}_{l}
\end{bmatrix}  
= 
\begin{bmatrix}
  \frac{d\hat{\bf}_{-q}}{d \bs} \\
 0
\end{bmatrix},  
\label{eq:k=-q_equation}
\end{equation} 
while for  $k=+q$,
\begin{equation}
i q\omega_0 \mathcal{I}_u
\begin{bmatrix}
  \hat{\bv}_{q} \\
 \hat{\eta}_{q}
\end{bmatrix}  
- \sum_{l= 0}^q T_{q-l}  
\begin{bmatrix}
  \hat{\bv}_{l} \\
 \hat{\eta}_{l}
\end{bmatrix}  
= 
\begin{bmatrix}
  \frac{d\hat{\bf}_{q}}{d \bs} \\
 0
\end{bmatrix}.
\label{eq:k=+q_equation}
\end{equation} 
Stacking the blocks together one below the other results in a linear system of equations that takes the form
\begin{equation}
\begin{bmatrix}
\mathcal{D} - \mathcal{T } (T_{m})
\end{bmatrix}
\widehat{\mathcal{V}} = \widehat{\mathcal{R}},
\label{n1}
\end{equation}
where $ \widehat{\mathcal{V}} = \left [ \widehat{\mathcal{V}}_{-q}, \dots, \widehat{\mathcal{V}}_{0}, \dots, \widehat{\mathcal{V}}_{q} \right ]^\top$ is the block vector of unknowns, with $\widehat{\mathcal{V}}_k = [\hat{\bv}_k, \hat{\eta}_k]^\top $. The right hand side is $\widehat{\mathcal{R}} = [ \widehat{\mathcal{R}}_{-q}, \dots, \widehat{\mathcal{R}}_{0}, \dots, \widehat{\mathcal{R}}_{q}]^\top$ with $\widehat{\mathcal{R}}_k = \left [ \hat{\frac{d \bf_k}{d \bs}}, 0 \right ]^\top$, while $\mathcal{D}$ is the block diagonal matrix $\mathcal{D}=diag[ \mathcal{D}_{-q}, \dots, \mathcal{D}_0, \dots, \mathcal{D}_q ]$, with $\mathcal{D}_k=i k \omega_0  \mathcal{I}_u$.  Matrix $\mathcal{T } (T_{m})$ has a block Toeplitz form, 
\begin{equation}
\mathcal{T } (T_{m}) = 
\begin{bmatrix}
T_0   &  T_{-1}  & \dots & T_{-q}  &   &  &   \\
T_1  &  T_{0}   & T_{-1} & \dots  & T_{-q} &  &  \\
     &  \ddots   & \ddots  & \ddots  &  \dots & \ddots &   \\
T_q   &  \dots   & T_{1}  & T_0   & T_{-1} & \dots & T_{-q}  \\
     &  \dots    & \ddots  & \ddots   & \ddots  & \ddots & \dots  \\
     &     & T_q  & \dots   & T_{1}  & T_{0}  & T_{-1}  \\
          &   &   & T_q  & \dots & T_1 & T_0  \\
\end{bmatrix},
\label{eq:Block_Toeplitz_matrix}
\end{equation}
with the same blocks in each  diagonal. 

System \eqref{n1} can be also written in expanded matrix form as
\begin{equation}
\begin{bmatrix}
\mathcal{D}_{-q}-T_0   &  -T_{-1}  & \dots & -T_{-q}  &   &  &   \\
-T_1  &  \mathcal{D}_{-q+1}-T_{0}   & -T_{-1} & \dots  & -T_{-q} &  &  \\
     &  \ddots   & \ddots  & \ddots  &  \dots & \ddots &   \\
-T_q   &  \dots   & -T_{1}  & \mathcal{D}_{0}-T_0   & -T_{-1} & \dots & -T_{-q} \\
     &  \dots    & \ddots  & \ddots   & \ddots  & \ddots & \dots  \\
     &     & -T_q  & \dots   & -T_{1}  & \mathcal{D}_{q-1}-T_{0}  & -T_{-1}  \\
          &   &   & -T_q  & \dots & -T_1 & \mathcal{D}_{q}-T_0  \\
\end{bmatrix}
\begin{bmatrix}
\widehat{\mathcal{V}}_{-q} \\ \widehat{\mathcal{V}}_{-q+1} \\ \vdots \\ \widehat{\mathcal{V}}_{0} \\ \vdots \\ \widehat{\mathcal{V}}_{q-1} \\\widehat{\mathcal{V}}_{q}
\end{bmatrix}
=
\begin{bmatrix}
\widehat{\mathcal{R}}_{-q} \\ \widehat{\mathcal{R}}_{-q+1} \\ \vdots \\ \widehat{\mathcal{R}}_{0} \\ \vdots \\ \widehat{\mathcal{R}}_{q-1} \\\widehat{\mathcal{R}}_{q}
\end{bmatrix}.
\label{eq:harmonic_balancing_system}
\end{equation}

The above matrix, known known as Hill matrix \cite{Lazarus_Thomas_2010, WereleyThesis}, contains square blocks with dimensions $(N_u+1)\times(N_u+1)$, and thus has very large storage requirements. The solution of system \eqref{n1} can be written symbolically as
\begin{equation}
\widehat{\mathcal{V}}=\mathcal{H} \widehat{\mathcal{R}},
\label{eq:symbolic_solution}
\end{equation}
where
\begin{equation}
\mathcal{H}=
\begin{bmatrix}
\mathcal{D} - \mathcal{T } (T_{m})
\end{bmatrix}^{-1},
\label{eq:shadowing_resolvent_operator}
\end{equation}
is the matrix that maps the input $\widehat{\mathcal{R}}$ to the output $\widehat{\mathcal{V}}$. As $q\to \infty$, $\mathcal{H}$ becomes an operator, termed here shadowing harmonic operator. Note that equations \eqref{eq:ODE_tangent_dilation} and \eqref{eq:orthogonality} form a linear, time-varying periodic system,
\begin{equation}
\begin{bmatrix}
 \mathcal{I}_{N_u} & 0 \\
 0                 & 0
\end{bmatrix}
\frac{d}{dt}
\begin{bmatrix}
 \bv \\
 \eta
\end{bmatrix}
=
\begin{bmatrix}
\frac{\partial \bf}{\partial \bu}  & \bf \\
 {\bf}^\top                        & 0  
\end{bmatrix}
\begin{bmatrix}
 \bv \\
 \eta
\end{bmatrix}
+
\begin{bmatrix}
\frac{\partial \bf}{\partial \bs}  \\
 0
\end{bmatrix}.
\label{eq:linear_periodic_system}
\end{equation}
Thus, the shadowing harmonic operator is identical to the standard harmonic operator, defined in \citep{WereleyThesis}, applied to the above system. The properties of this operator will be examined in the next section for the Kuramoto-Sivashinsky equation. 

The sensitivity $\frac{d \overline{J}}{ds}$ is computed from \cite{WANG2014210},
\begin{equation}
\frac{d \overline{J}}{ds} = \frac{1}{T} \int_0^T \frac{\partial J}{ \partial \bu} \bv + \frac{\partial J}{ \partial \bs} + \eta(t) (J(\bu; \bs; t)-\overline{J}) dt,
\label{eq:sensitivity_time_domain}
\end{equation} 
which in the frequency domain can be written as,
\begin{equation}
\frac{d \overline{J}}{ds} = \left(\widehat{\frac{\partial J}{ \partial \bs}}\right)_0 + \sum_{k=-q}^{q} \left( \widehat{\frac{\partial J}{ \partial \bu}}\right)_{-k} \hat{\bv}_k +  \sum_{\substack{k=-q \\ k\ne 0}}^q \hat{\eta}_{-k} \hat{J}_k
\label{eq:sensitivity_freq_domain}
\end{equation} 

In the above formulation, the reference trajectory $\bu(t; \bs)$ was expanded using the same number of Fourier modes as the solution, $2q+1$, and each diagonal of the block Toeplitz matrix $\mathcal{T } (T_{m})$ corresponds to one harmonic of $\bu(t; \bs)$. This is however not necessary. For example, if the spectral content of $\bu(t; \bs)$ is concentrated in a few frequencies, then only the relevant diagonals of $\mathcal{T } (T_{m})$ need to be retained. This can lead to enormous savings in the storage requirements and solution time of system \eqref{eq:harmonic_balancing_system}. In the limiting case, where only the time-average of $\bu(t;\bs)$ is retained, the equations decouple and the $k$-th component $\widehat{\mathcal{V}}_k = [\hat{\bv}_k, \hat{\eta}_k]^\top $ can be obtained from 
\begin{equation}
\left(\mathcal{D}_{k}-T_0 \right) \widehat{\mathcal{V}}_k=\widehat{\mathcal{R}}_{k} 
\end{equation}
In this case, the harmonic balancing method becomes identical to the standard resolvent analysis \cite{mckeon_sharma_2010}.

Some comments are warranted here to clarify an underlying assumption of the above formulation. Suppose that the reference trajectory $\bu(t; \bs)$, and thus  $\frac{\partial \bf}{\partial \bu}(t)$ and $\bf(t)$, are exactly periodic with period $T$ and the minimisation problem \eqref{eq:minimisation_3} is solved in the time domain using the multiple shooting shadowing method, \cite{bloniganMSS,Shawki_Papadakis_2019}. If the trajectory is sampled at points $t_0 \dots t_K$ during $T$, then the solution $\bv(t_i), \eta(t_i)$ is sought at $K+1$ points  $i=0\dots K$, thus there are $2(K+1)$ unknowns. There are $K$ intervals, thus \eqref{eq:ODE_tangent_dilation} provides $K$ equations, while \eqref{eq:orthogonality} provides additional  $K+1$ equations. The remaining equation arises from the solution of the minimisation problem  \eqref{eq:cost_function_3}. The solution of this problem will not necessarily yield a periodic solution, i.e. $\bv(t_0)$ will not necessarily be equal to $\bv(t_K)$.

In the above formulation, we have not explicitly considered the minimisation of the cost function \eqref{eq:cost_function_3}; instead we closed the system assuming periodicity, i.e. $\bv(t_0)=\bv(t_K)$ and $\eta(t_0)=\eta(t_K)$. The same assumption is made in the periodic shadowing method of Lasagna  \citep{LASAGNA2019119}, and leads to a sensitivity error that initially decays at a rate $1/T$, followed by the asymptotic rate $1/\sqrt{T}$ (the latter dictated by the central limit theorem). There is however an important difference compared to the present method: In \citep{LASAGNA2019119}, the time transformation $\tau(t)$ is linear with respect to $t$, leading to a constant $\eta(t)$. 
In the present method, we do not prescribe any form of $\tau(t)$, so $\eta(t)$ is unknown and is obtained by imposing the orthogonality constraint \eqref{eq:orthogonality} at every point along the  trajectory. 
Another difference is that our method is formulated in the frequency domain instead of the time domain. As will be seen later, this can lead to significant simplifications, and allows one to obtain deep physical insight on the dominant factors that determine the sensitivity, which is not possible in the time domain.

An approach closely related to the present one was proposed recently by Padovan et al \cite{padovan_otto_rowley_2020}. The authors perform a frequency-domain analysis of periodic perturbations about a periodically time-varying base flow. If the base flow satisfies the governing equations, the perturbations are governed by $\frac{d \boldsymbol{q}'}{dt} = \frac{\partial \bf}{\partial \bu} \boldsymbol{q}'\ +  \boldsymbol{g}'$, which is similar to \eqref{eq:ODE_tangent_dilation}. The authors restrict the harmonic operator to a subspace that is orthogonal to the direction of the phase shift given by $\hat{\bf}_k$. This is achieved by projecting out of  $\hat{\boldsymbol{g}}'_k$ (the Fourier coefficients of the forcing $\boldsymbol{g}'(t)$) the component that would lead to a non-zero projection of the solution $\hat{\boldsymbol{q}}'_k$ to $\hat{\bf}_k$. 
In the present paper, we seek the sensitivity with respect to $\bs$, thus the forcing vector $\boldsymbol{g}'(t)$ takes the particular form $\boldsymbol{g}'(t)=\frac{\partial \bf}{\partial \bs}(t)$. This vector is subsequently modified by adding $\eta(t)\bf(t)$, where the time dilation $\eta(t)$ is computed so that the orthogonality constraint is satisfied at all time instants, as already mentioned. 


In the above frequency-domain formulation, we seek a harmonic solution that remains bounded, see expansions \eqref{Fourier Series 1}, and does nor suffer from exponentially growing terms. There is a price to pay however; the frequencies are all coupled together leading to large storage requirements for the Hill matrix, see \eqref{eq:harmonic_balancing_system}. The properties of the shadowing harmonic operator will be examined in the next section for the Kuramoto-Sivashinsky equation. This test case is small enough that the Hill matrix can be stored and the linear system \eqref{eq:harmonic_balancing_system} is solved directly with LU decomposition. In section \ref{Simplified Fourier} we propose an iterative method that mitigates the storage and solution time requirements. 

\section{Application to the Kuramoto-Sivashinsky equation}\label{sec:app_Kuramoto_Sivashinsky}
We apply the method proposed in the previous section to  the Kuramoto Sivashinsky (KS) equation, which displays complex spatio-temporal chaos and is frequently used in the literature as a test case for chaotic systems \cite{HYMAN1986113}. The equation takes the form 
\begin{equation}
\begin{aligned}
&\frac{\partial u}{\partial t}=-\left [ (u+c) \frac{\partial u}{\partial x} + \frac{\partial^2 u}{\partial x^2}+ \frac{\partial^4 u}{\partial x^4} \right ] \mbox{ with }x\in[0,L],\\
& u(0,t)=u(L,t)=0, \\
& \frac{\partial u}{\partial x} (0,t) =\frac{\partial u}{\partial x} (L,t)=0,
\end{aligned}
\label{KS eqn}
\end{equation}
where $c$ is an artificially introduced parameter \cite{BLONIGAN201416}. The term $ \frac{\partial^2 u}{\partial x^2}$ is responsible for energy production, while $\frac{\partial^4 u}{\partial x^4}$ adds dissipation to the system. We set $L = 128$ to generate chaotic solutions \cite{HYMAN1986113} and discretise \eqref{KS eqn} with a second order finite difference scheme with $\delta x=1$. For $c=0$, the dynamical system has $16$ positive Lyapunov exponents, the largest of which is $\lambda_1 = 0.093$ \cite{BLONIGAN201416}.  

Two objective functions are considered, the space-time average of the state $u(x,t;c)$
\begin{equation}
\label{KS objective u}
\overline{J_1}(c) = \frac{1}{TL} \int_0^T \int_0^L u(x,t;c) dx \, dt =\frac{1}{L} \int_0^L \overline{u}(x;c) dx  
\end{equation}
and of the total kinetic energy  
\begin{equation}
\label{KS objective 2}
\overline{J_2}(c) = \frac{1}{TL} \int_0^T \int_0^L u^2(x,t;c) dx \, dt= \frac{1}{L} \int_0^L \overline{u^2}(x;c) dx = 
\frac{1}{L} \int_0^L \left (\overline{u}^2+ \overline{u'^2} \right) \, dx
\end{equation}
where an overbar $\overline{()}$ denotes time-average and a prime  ${()'}$ the fluctuation around the average, i.e.\ $u=\overline{u}+u'$. We seek their sensitivities with respect to $c$, i.e.\ $\frac{d\overline{J_1}(c)}{dc}$ and  $\frac{d\overline{J_2}(c)}{dc}$, thus $s=c$. 
 
\begin{figure}[!htb]
\centering
\begin{subfigure}[b]{0.49\textwidth} 	\includegraphics[scale=0.40, clip]{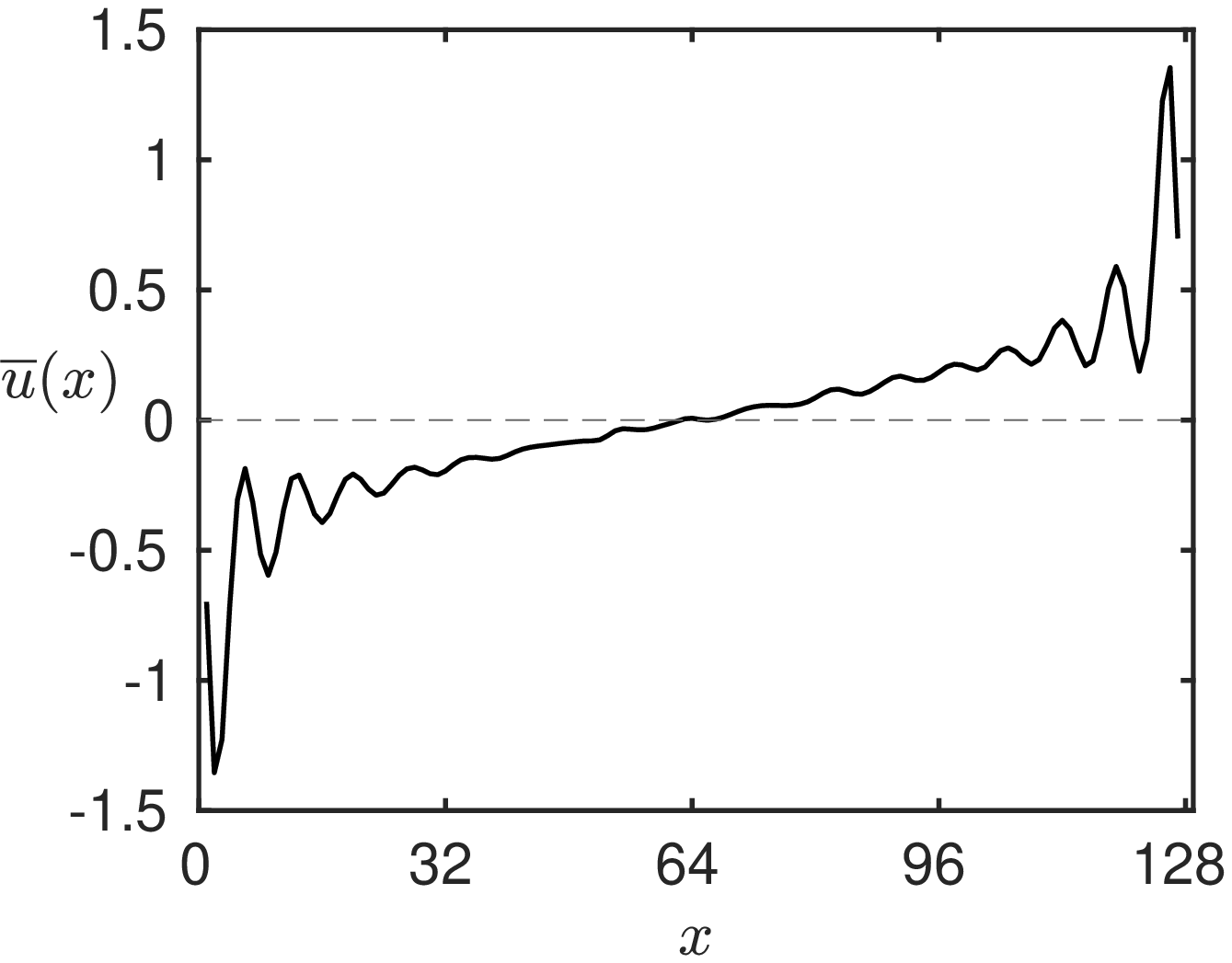}
\caption{}
\label{fig:KS_TA_u}
\end{subfigure}
\begin{subfigure}[b]{0.49\textwidth} 	\includegraphics[scale=0.40, clip]{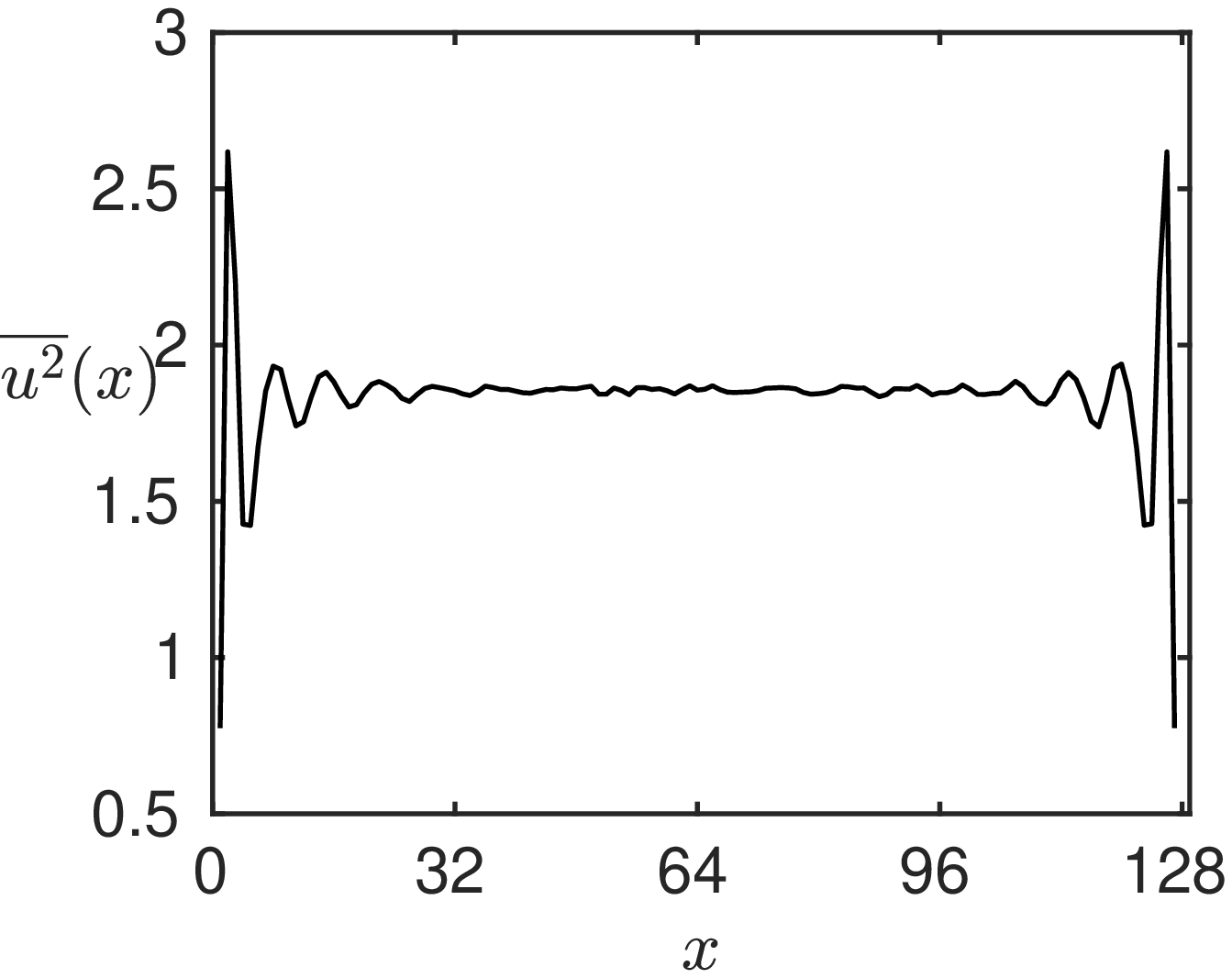}
\caption{}
\label{fig:KS_TA_u2}
\end{subfigure}
\begin{subfigure}[b]{0.49\textwidth} 	\includegraphics[scale=0.40, clip]{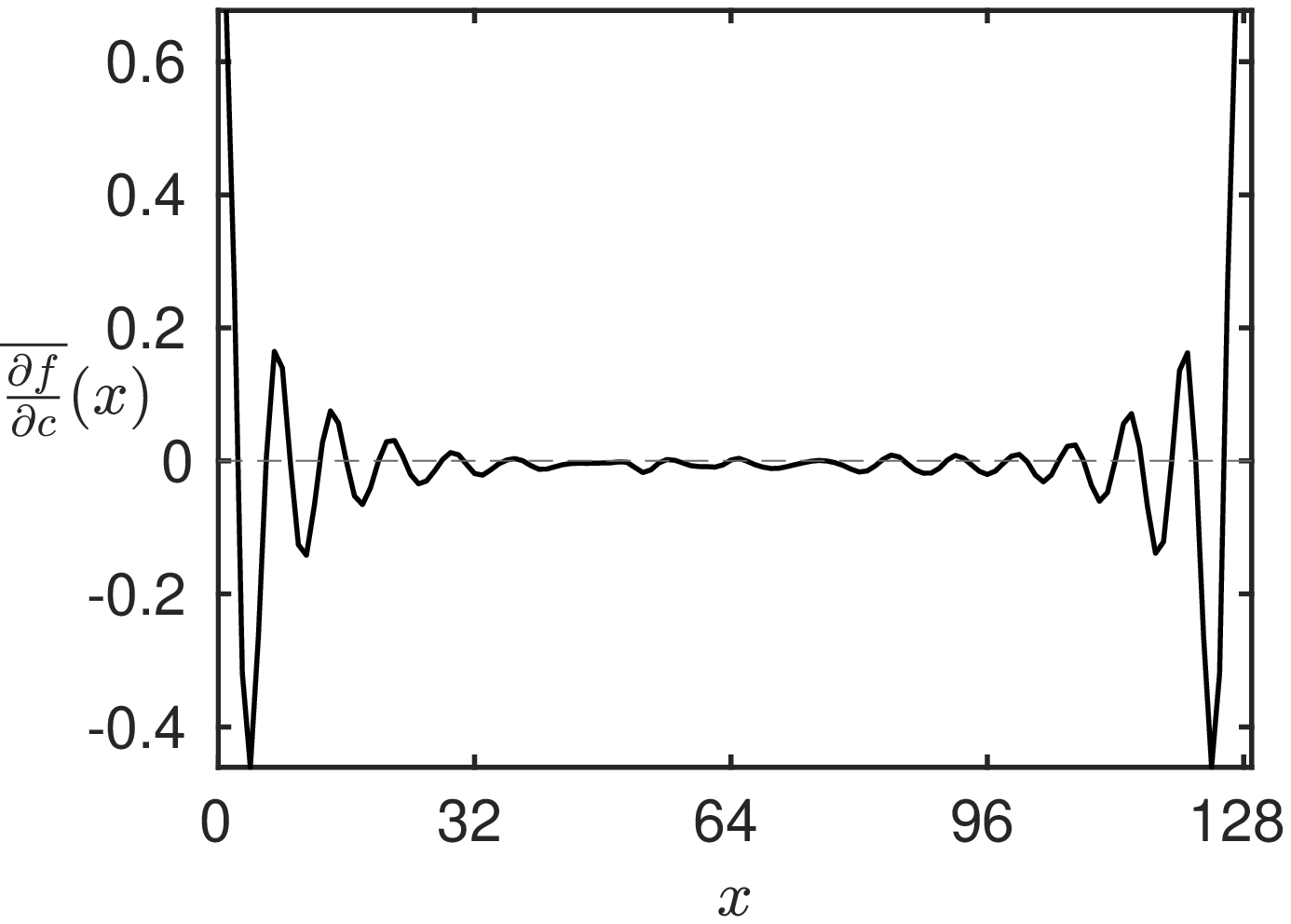}
\caption{}
\label{fig:KS_TA_dfds}
\end{subfigure}
\caption{Spatial distribution of the time-averages (a) $\overline{u}(x)$, (b) $\overline{u^2}(x)$ and (c) $\overline {\frac{df}{dc}}(x)$ for $c = 0$ and $T = 100$. The results are averaged over 1000 random initial conditions.}
\label{KS_time_averages}
\end{figure} 

The time-averages $\overline{u}(x)$,  $\overline{u^2}(x)$ and $\overline{\frac{df}{dc}}(x)$ for $c=0$ are plotted against $x$ in figure \ref{KS_time_averages}. It can be seen that $\overline{u}(x)$, shown in panel (a), takes both positive and negative values, so it is not intuitively evident which part of the domain contributes most to the integral $\overline{J_1}(c)$ and its sensitivity. The kinetic energy $\overline{u^2}(x)$, shown in panel (b), has an almost symmetric shape with peaks close to the boundaries, but remains flat in the middle of the domain. In the latter region, $\overline{u}(x)$ attains low values, so the total kinetic energy consists mainly of the fluctuation energy,  $\overline{u'^2}$. Finally, the time-average forcing term $\overline{\frac{df}{dc}}(x)$, shown in panel (c), fluctuates around 0 close to the boundaries and is almost negligible elsewhere. Again, it is not trivial to foresee how this term will affect the sensitivities of $\overline{J_1}(c)$ and $\overline{J_2}(c)$ with respect to $c$.

\begin{figure}[!htb]
\centering
\includegraphics[scale=0.60, clip]{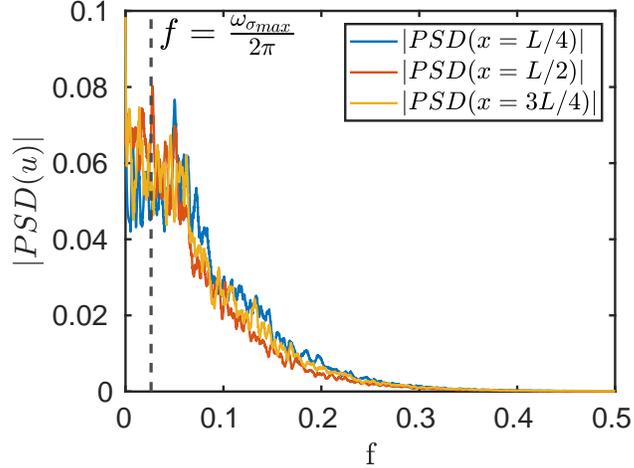}
\caption{Spectra of $u(x,t)$ at  $x = \frac{L}{4}$, $\frac{L}{2}$ and $\frac{3L}{4}$ for $c=0$ obtained from a trajectory with length $T = 10000$. The spectra were smoothed with a $5-th$ order Savitzky-Golay convolution filter with 5 averaging windows. The vertical dashed line indicates the angular frequency $\omega \approx 0.164$ that gives the maximum gain $\sigma(\omega)$ in figure \ref{fig:KS_singular_values}.}
\label{fig: FFT Spectra}
\end{figure}

The spectra of $u(x,t)$ at three locations for $c = 0$ are shown in figure \ref{fig: FFT Spectra}. The plot indicates that the solution has a spectral footprint in the frequency range $[0.,0.3]$, with the highest energy content in the region $[0,0.07]$. This information will prove useful later in the paper.

From equation \eqref{eq:sensitivity_freq_domain}, we have
\begin{equation}
\frac{d \overline{J_1}}{dc} = \frac{1}{L} \int_0^L  \left(\hat{\bv}_0 +  \sum_{\substack{k=-q \\ k\ne 0}}^q \hat{\eta}_{-k} \hat{\bu}_k\right) \, dx,
\label{eq:sensitivity_J1_freq}
\end{equation}
and
\begin{equation}
\frac{d \overline{J_2}}{dc} = \frac{1}{L} \int_0^L  \left(2 \sum_{k=-q}^{q}\bu_{-k} \hat{\bv}_k + \sum_{\substack{k=-q \\ k\ne 0}}^q \hat{\eta}_{-k} \widehat{\bu^2}_k \right) \, dx.
\label{eq:sensitivity_J2_freq}
\end{equation}
The above analytical expressions can be interpreted physically. The distribution of the integrands over $x$ reveals the parts of the domain that mostly affect the quantity of interest; this information can be useful for control applications for example.  

\begin{figure}[!htb]
\centering
\begin{subfigure}[b]{0.49\textwidth} 	\includegraphics[scale=0.40, clip]{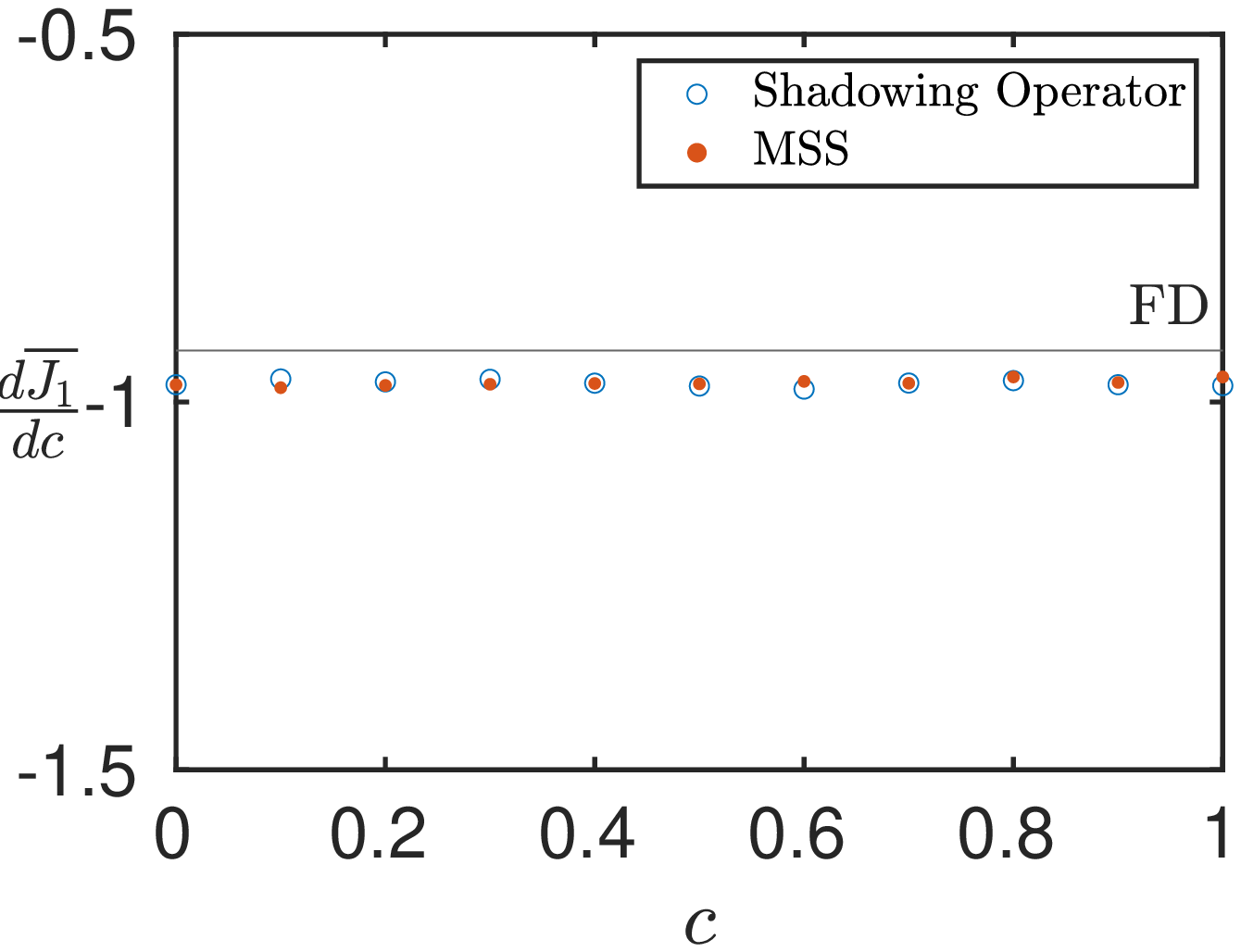}
\caption{}
\label{fig: Full MSS FD Comparison u}
\end{subfigure}
\begin{subfigure}[b]{0.49\textwidth} 	\includegraphics[scale=0.40, clip]{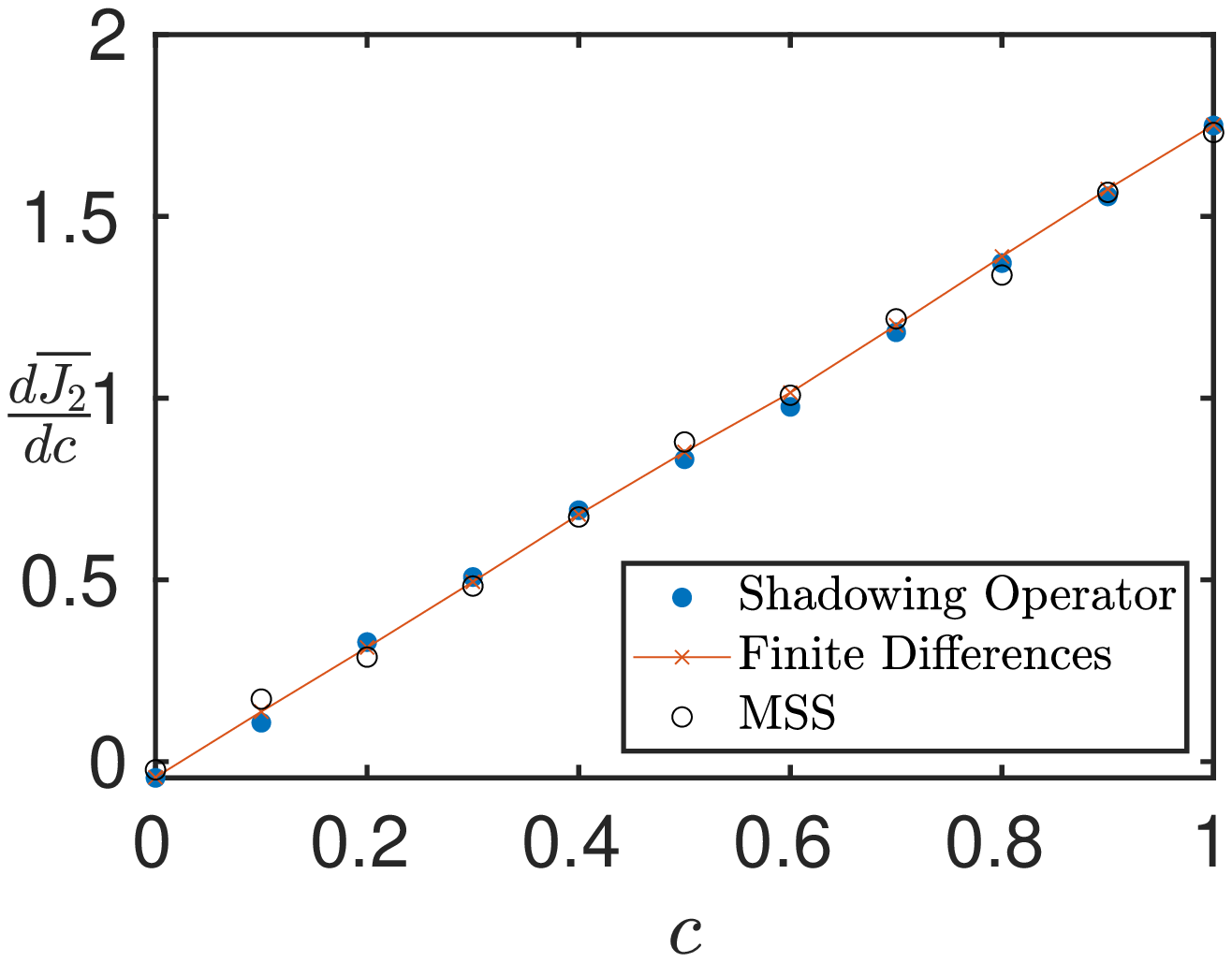}
\caption{}
\label{fig: Full MSS FD Comparison u2}
\end{subfigure}
\caption{Sensitivities $\frac{d\overline{J_1}}{dc}$ (a) and $\frac{d\overline{J_2}}{dc}$ (b) for various values of parameter $c$, obtained with $T = 100$ and $f \in [-0.3,0.3]$. The finite difference (FD) data were taken from \cite{BLONIGAN201416}. The MSS values were obtained using $K = 10$ time segments of length $\Delta T=10$. Values were averaged over 100 random initial conditions.}
\label{fig: Full MSS FD Comparison}
\end{figure}

As can be seen from figure \ref{fig: Full MSS FD Comparison}, the sensitivities obtained using the shadowing harmonic operator match very well with those obtained using the preconditioned MSS \cite{Shawki_Papadakis_2019}. A comparison with finite difference (FD) data is also presented. For $\frac{d \overline{J_1}}{dc}$, there is a small bias, which has also been observed in the time-domain formulation of the method \cite{Shawki_Papadakis_2019,Kantarakias_Shawki_Papadakis_2020,BLONIGAN201416}; for $\frac{d \overline{J_2}}{dc}$ the matching with FD is very good.

\begin{figure}[!htb]
	\centering
	\includegraphics[scale=0.45, clip]{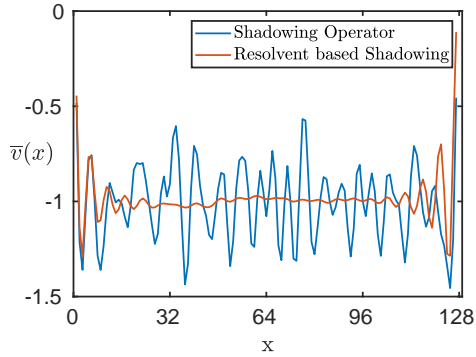}
	\caption{Comparison of $\overline{\bv}(x)=\hat{\bv}_0(x)$ between the harmonic shadowing operator and the resolvent-based approximation (described in section \ref{Simplified Fourier}), for $c = 0$ and $T = 100$, averaged over 100 random initial conditions. } 
	\label{fig: Comparison Full Decoupled vTA}
\end{figure}

Figure \ref{fig: Comparison Full Decoupled vTA} shows the variation of $\overline{\bv}(x)=\hat{\bv}_0(x)$ against $x$. The values fluctuate around $-1$ almost uniformly in the whole domain, thus changing $c$ to $c+\delta c$ reduces $\overline{u}(x)$ everywhere by the same amount on average.   

\begin{figure}[!htb]
	\centering
	\includegraphics[scale=0.450, clip]{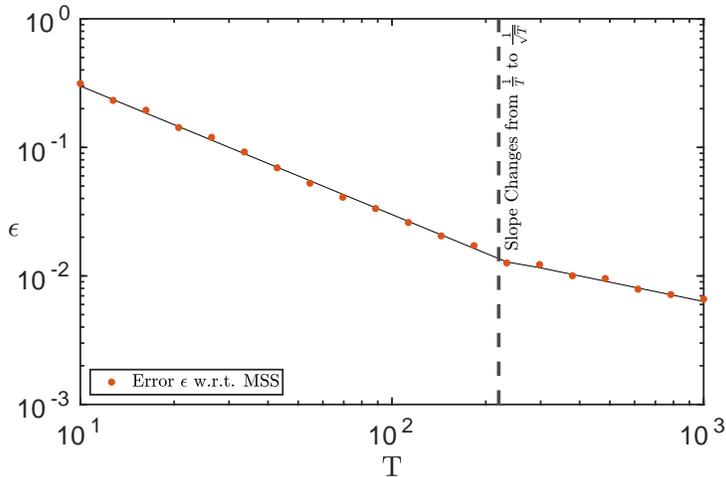}
	\caption{Normalised difference in sensitivity $\frac{d \overline{J_1}}{dc}$ as computed by the shadowing harmonic method and MSS. The vertical dashed line marks the value of $T$ at which the decay rate changes from $\frac{1}{T}$ to $\frac{1}{\sqrt{T}}$. } 
	\label{fig: Full Error Convergence}
\end{figure}

In figure \ref{fig: Full Error Convergence}, we plot the normalised difference in the sensitivity $\frac{d \overline{J_1}}{dc}$ from MSS and the shadowing harmonic method, $\epsilon = \frac{\left(\frac{d \overline{J_1}}{dc}\right)_{SH} - \left(\frac{d \overline{J_1}}{dc}\right)_{MSS}}{\left(\frac{d \overline{J_1}}{dc}\right)_{MSS}}$, against $T$. The difference initially decays at a rate $\frac{1}{T}$ (similarly to \cite{LASAGNA2019119}) and then at  $\frac{1}{\sqrt{T}}$ (as dictated by the central limit theorem). The frequencies considered are $f \in [-0.3,0.3]$, for $c = 0$. Each point was obtained by averaging over a large number of initial conditions dictated by the value of $T$ and enough for the sensitivity to converge to two significant digits. 

Evaluating the sensitivities using the preconditioned MSS with $K = 10$ segments requires approximately 55s in a 3.2 GHz Intel i7-8700 CPU. On the other hand, the shadowing harmonic method requires an order of magnitude less time (approximately 3s) and this cost is independent of the number of positive Lyapunov exponents. This comparison however is case dependent. The storage requirements are much larger, while the cost of the LU decomposition of the block-diagonal system \eqref{eq:harmonic_balancing_system} scales linearly with the number of diagonals and the cost of inversion of a single block, which scales as $\sim (N_u+1)^3$ for dense blocks. Thus for larger systems, both storage and solution costs grow fast.

Below we investigate in more detail the properties of the shadowing harmonic operator, while in the next section \ref{Simplified Fourier} we explore an approach that can mitigate the aforementioned rapid growth of computational cost and storage requirements for larger systems.

\subsection{Singular value decomposition of the shadowing harmonic operator}
The singular values $\sigma_i$ of the shadowing resolvent matrix $\mathcal{H}$, defined in \eqref{eq:shadowing_resolvent_operator}, are obtained from the solution of the  eigenvalue problem,
\begin{equation}
\mathcal{H}^* \mathcal{H} \bphi = \sigma^2 \bphi.
\label{Resolvent Eigenvalue Problem FULL}
\end{equation}
The solution maximises the system gain, defined as the ratio of the (squared) 2-norm of the output (response $\widehat{\mathcal{V}}$) to that of the input (forcing $\widehat{\mathcal{R}}$), i.e.\
\begin{equation}
\label{Transfer Function FULL}
\sigma^2 =\max_{\widehat{\mathcal{R}}} \frac{\| \widehat{\mathcal{V}} \|_2^2}{\| \widehat{\mathcal{R}} \|_2^2},
\end{equation}
where $\| \widehat{\mathcal{V}}\|_2^2=\widehat{\mathcal{V}}^*\widehat{\mathcal{V}}$ and $\| \widehat{\mathcal{R}}\|_2^2=\widehat{\mathcal{R}}^*\widehat{\mathcal{R}}$. Since $\delta x=1$, the norms represent the discrete values of the integrals over the domain, for example $\| \widehat{\mathcal{V}}\|_2^2=  \widehat{\mathcal{V}}^*\widehat{\mathcal{V}}=  \sum_{k=-q}^{q}\hat{\bv}_k^* \hat{\bv}_k + \hat{\eta}_k^* \hat{\eta}_k  =
 \sum_{k=-q}^{q}\hat{\bv}_{-k} \hat{\bv}_k + \hat{\eta}_{-k} \hat{\eta}_k  \approx
\int_0^L \left(\overline{\bv(t)^2} +\overline{\eta(t)^2}\right) dx$. The first term represents physically (twice) the time-average kinetic energy of the response integrated over the domain. We could have defined a weighted 2-norm that eliminates the presence of $\overline{\eta(t)^2}$ from the integrand, but we did not pursue this. Other values of $\delta x$ could have been easily accounted for (again by defining an appropriately weighted 2-norm). The eigenvectors $\bphi_i$ are the right singular vectors of $\mathcal{H}$, and represent the optimal forcings, while the corresponding responses are the left singular vectors $\bpsi_i=\sigma_i^{-1}  \mathcal{H} \bphi_i$. 

\begin{figure}[!htb]
	\centering
	\includegraphics[scale=0.65, clip]{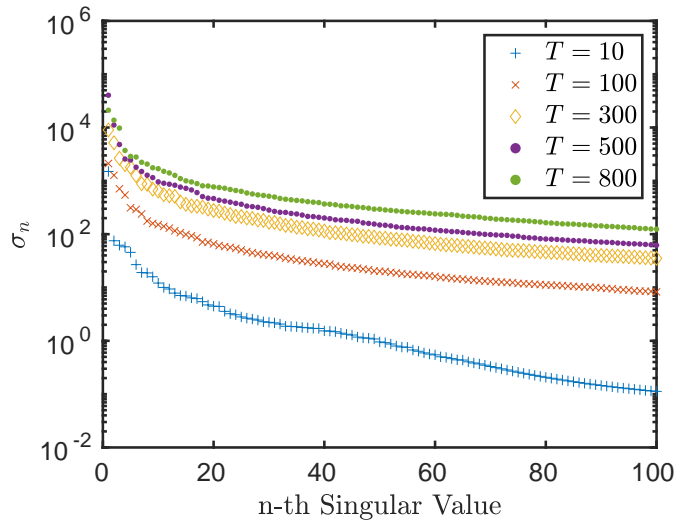}
	\caption{Leading singular values of  $\mathcal{H}$ for different values of $T$ and $c = 0$. The value of $q$ varies with $T$ such that the resolved frequencies are always in the interval $[-0.3,0.3]$.} 
	\label{SVD: Ranked Singular Values}
\end{figure}

Figure \ref{SVD: Ranked Singular Values} shows the 100 leading singular values for different $T$. Note that the largest singular value of $\mathcal{H}$, $\sigma_{max}^\mathcal{H}$, grows as $T$ increases, thus the smallest singular value of $\mathcal{H}^{-1}=\mathcal{D} - \mathcal{T } (T_{m})$,  $\sigma_{min}^{\mathcal{H}^{-1}}=\frac{1}{\sigma_{max}^\mathcal{H}}$, is reduced. The same behaviour is observed when the problem is formulated in the time domain \cite{bloniganMSS} and it is related to the lack of uniform hyperbolicity of the system. For such systems, the covariant Lyapunov vectors (CLVs) align at different locations along the trajectory and they are no longer linearly independent (this is equivalent to the eigenvectors of a matrix becoming parallel). This leads to local tangencies that result in the reduction of  $\sigma_{min}^{\mathcal{H}^{-1}}$ at $T$ increases.  On the other hand, for uniform hyperbolic systems, the angle between CLVs stays away from 0 and $\sigma_{min}^{\mathcal{H}^{-1}}$  remains bounded. For a discussion of the angles between CLVs for the turbulent flow around a cylinder see \cite{ni_2019}. Note also that as $T$ grows, apart from the largest singular values, the rest start to converge. This indicates that they represent the true behaviour of the system, i.e.\ they are physically meaningful.   

\begin{figure}[!htb]
\centering
\begin{subfigure}[b]{0.49\textwidth} 	\includegraphics[scale=0.40, clip]{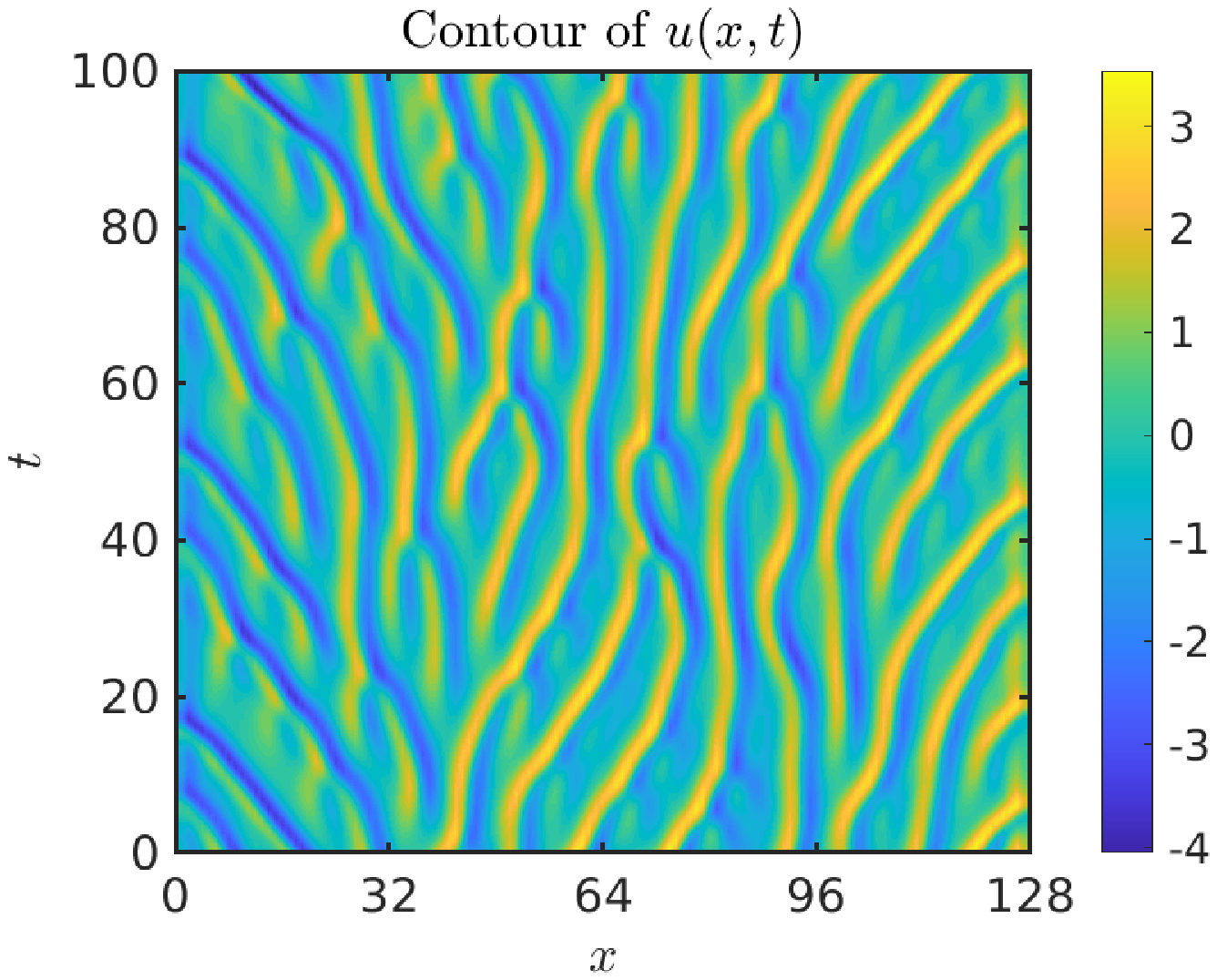}
\caption{}
\label{fig:u_of_t_spatiotemporal}
\end{subfigure}
\begin{subfigure}[b]{0.49\textwidth} 	\includegraphics[scale=0.40, clip]{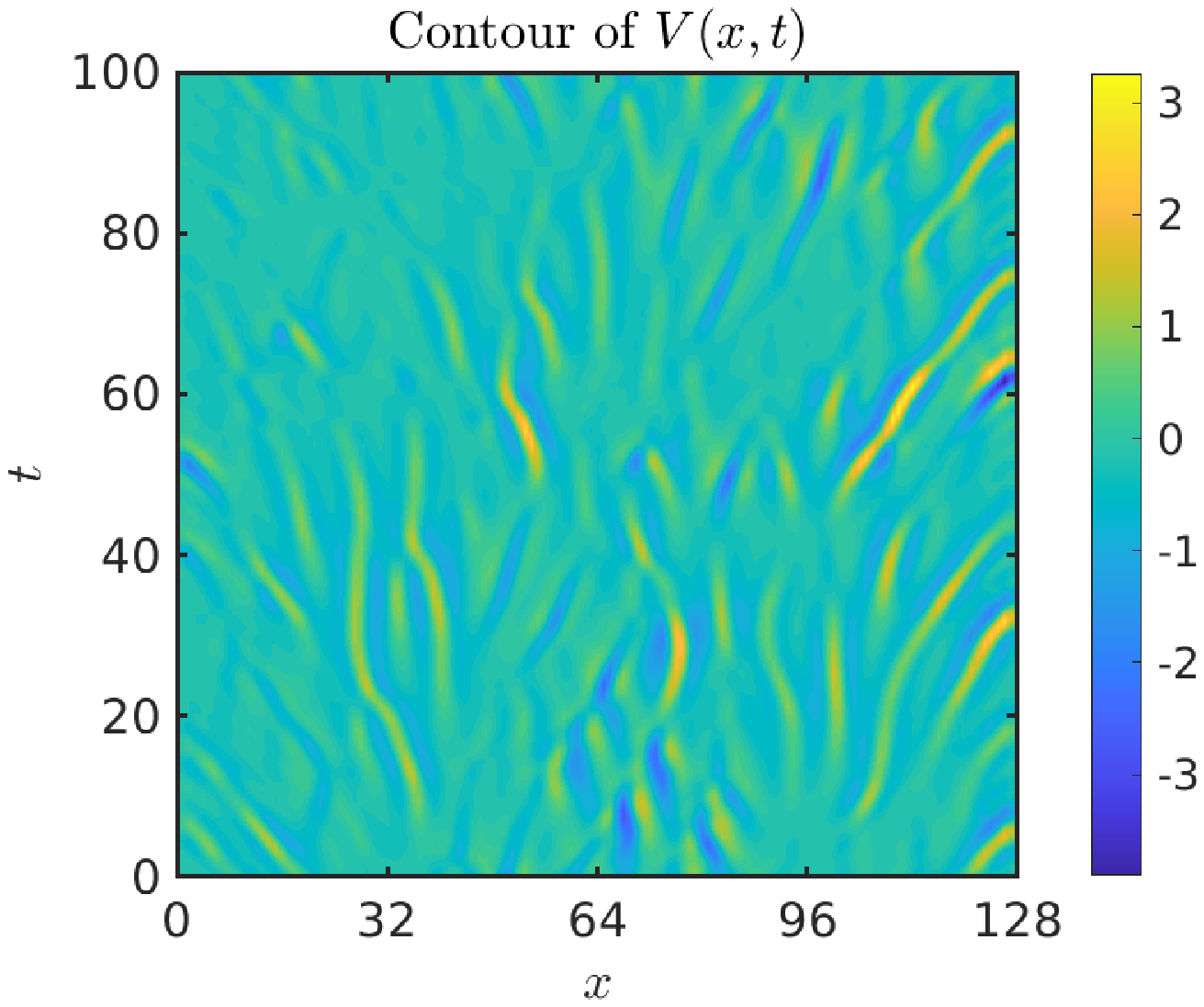}
\caption{}
\label{fig:v_of_t_spatiotemporal}
\end{subfigure}
\begin{subfigure}[b]{0.49\textwidth} 	\includegraphics[scale=0.40, clip]{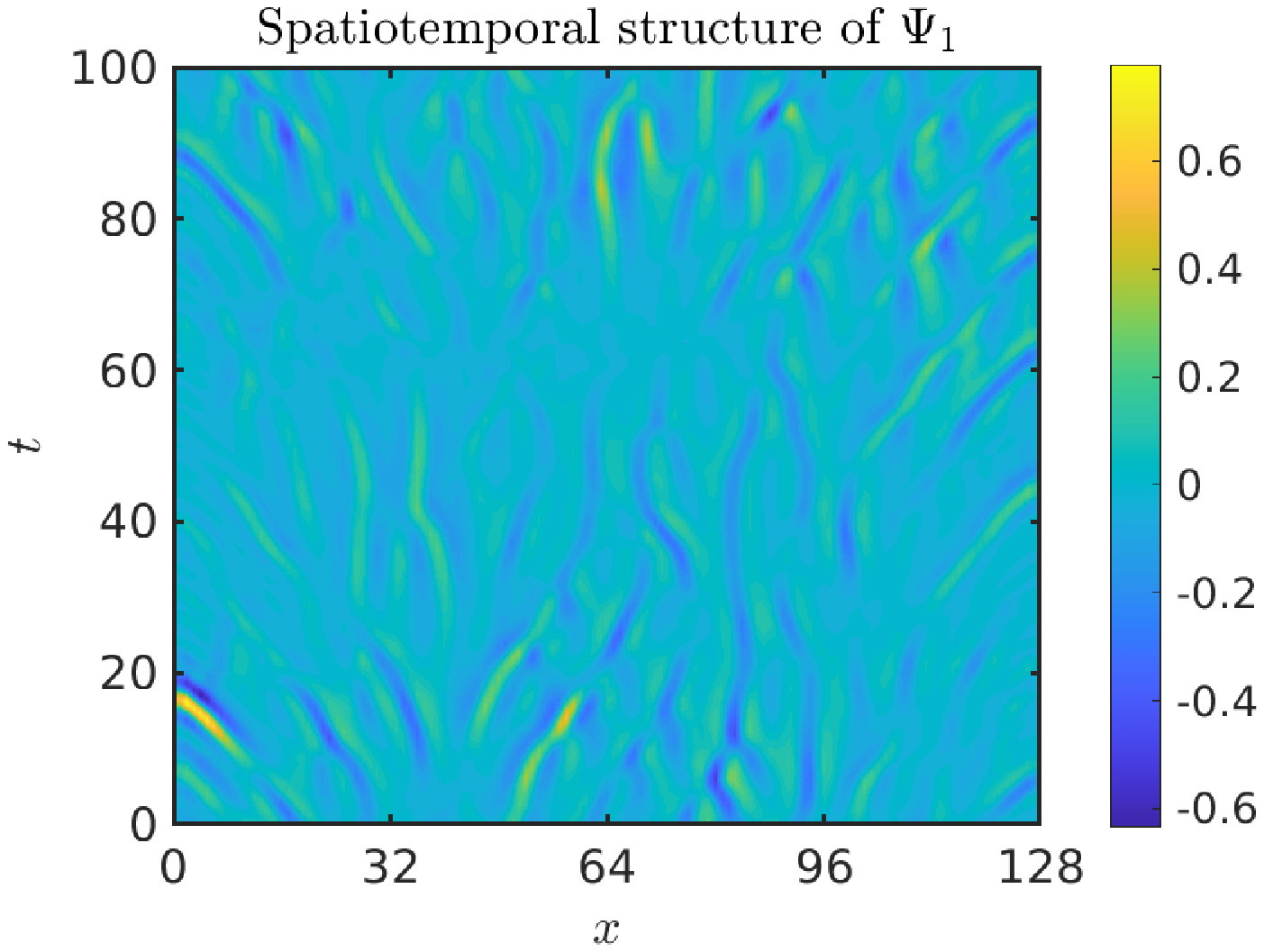}
\caption{}
\label{fig:Spatiotemporal_V1}
\end{subfigure}
\begin{subfigure}[b]{0.49\textwidth} 	\includegraphics[scale=0.40, clip]{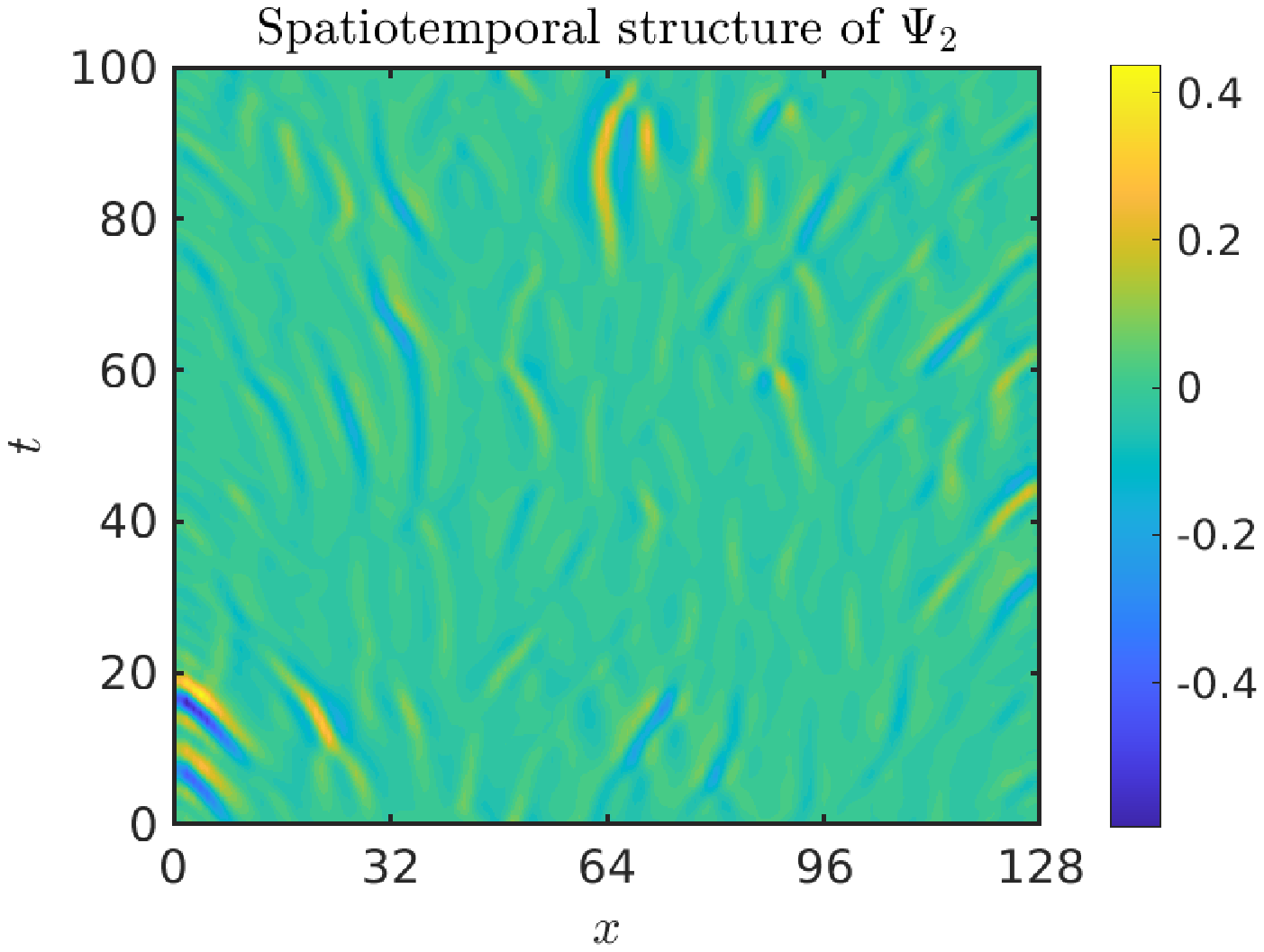}
\caption{}
\label{fig:Spatiotemporal_V2}
\end{subfigure}
\caption{Contours of (a) $\bu(x,t)$, (b) the sensitivity $\bv(x,t)$, (c) the first $\bpsi_1$ and (d) the second $\bpsi_2$ optimal responses in the $x-t$ plane. Results are shown for $T = 100$, $c = 0$ and $f \in [-0.3,0.3]$. }
\label{fig:Spatiotemporal Singular Vectors Responses}
\end{figure}

Contour plots of $u(x,t)$ in the $x-t$ plane for one realisation (with $c = 0$ and $T = 100$) are shown in figure  \ref{fig:u_of_t_spatiotemporal}.  Note the oscillatory back and forth motion in the middle region of the domain $[L/4,3L/4]$ that is captured in the spectra of figure \ref{fig: FFT Spectra}. The sensitivity $v(x,t)$ (obtained with $q = 30$) for the same realisation is shown in \ref{fig:v_of_t_spatiotemporal}. It follows a pattern similar to $u(x,t)$, but it is spotty (note the highly localised large  positive and negative values), probably due to the aforementioned tangencies. Figures \ref{fig:Spatiotemporal_V1} and \ref{fig:Spatiotemporal_V2} show contours of the spatio-temporal distribution of the optimal responses corresponding to the first (i.e.\ largest) and second  singular values, i.e.\ $\bpsi_1$ and $\bpsi_2$. The maps show again a wavy behaviour and are also locally spotty.  

\begin{figure}[!htb]
\centering
\begin{subfigure}[b]{0.49\textwidth} 	\includegraphics[scale=0.40, clip]{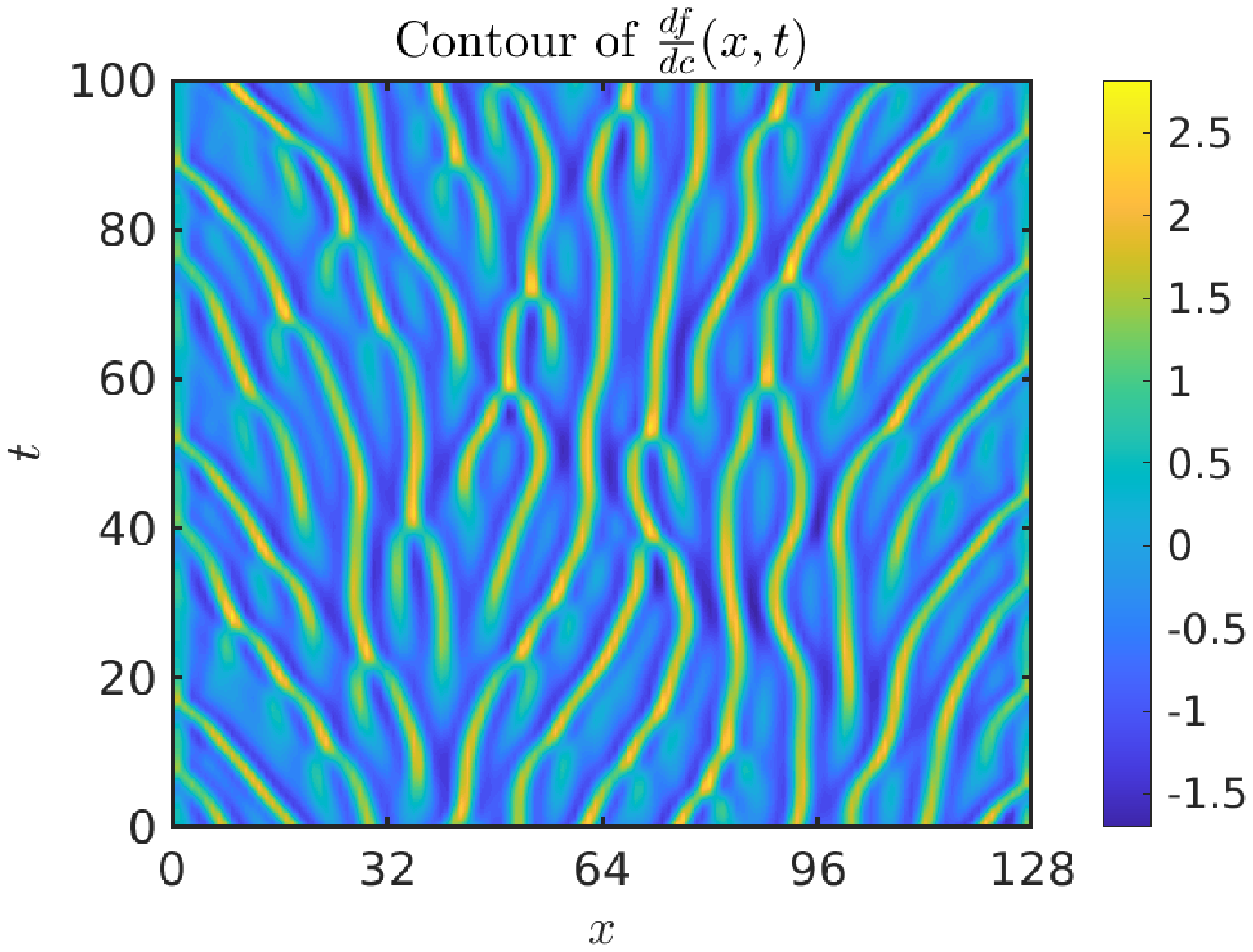}
\caption{}
\label{fig:dfds_of_t_spatiotemporal}
\end{subfigure}
\begin{subfigure}[b]{0.49\textwidth} 	\includegraphics[scale=0.40, clip]{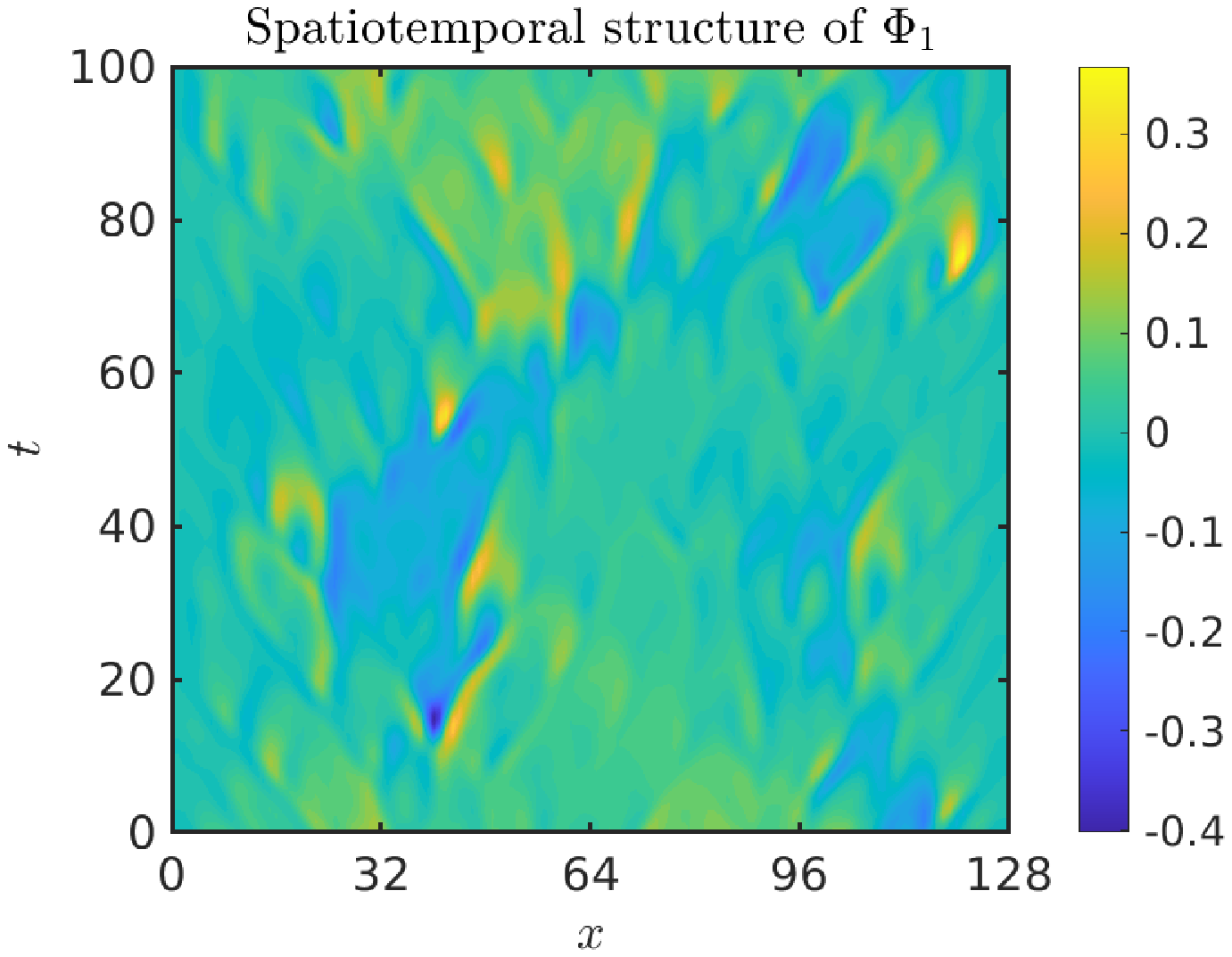}
\caption{}
\label{fig:Spatiotemporal_PHI1}
\end{subfigure}
\begin{subfigure}[b]{0.49\textwidth} 	\includegraphics[scale=0.40, clip]{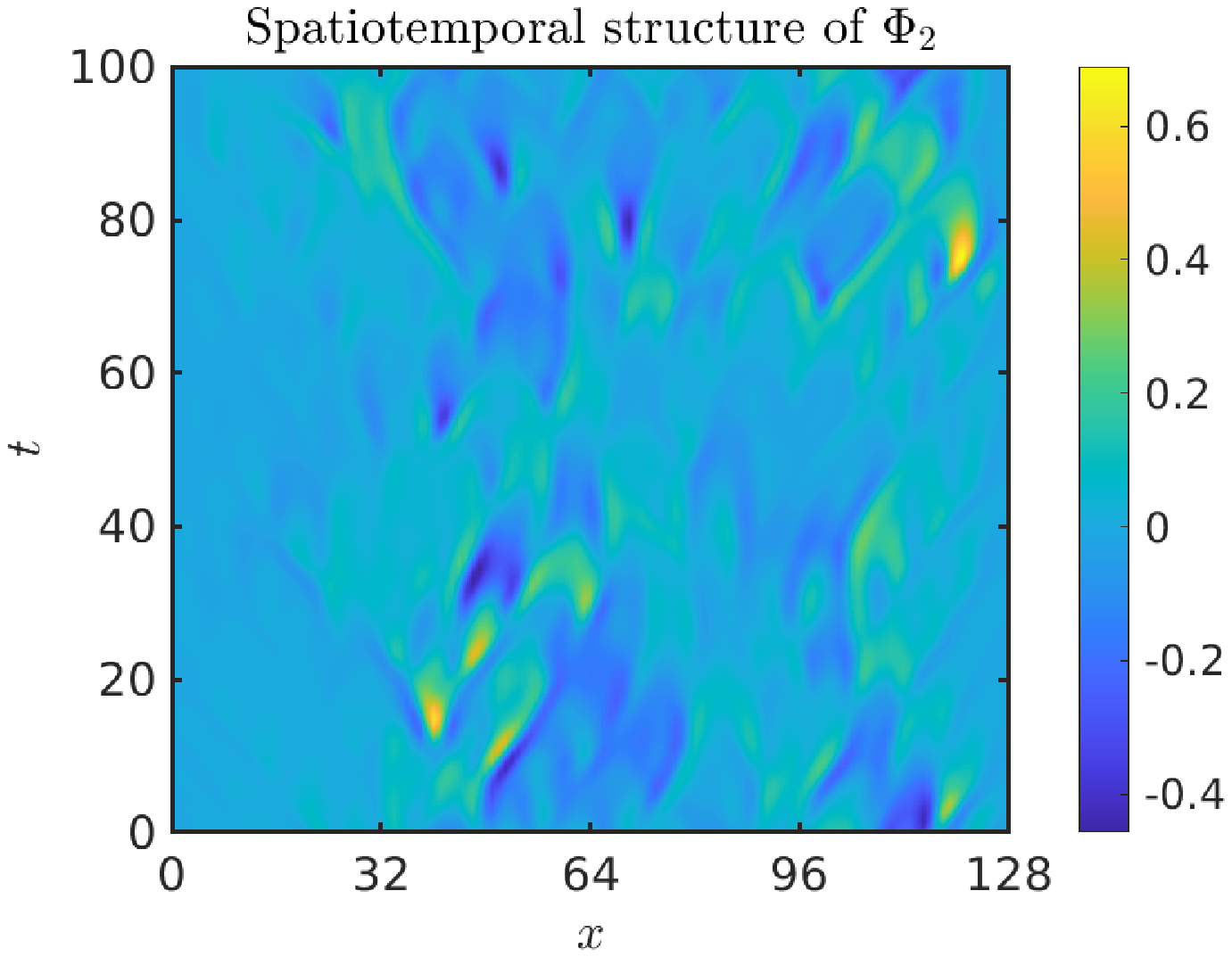}
\caption{}
\label{fig:Spatiotemporal_PHI2}
\end{subfigure}
\begin{subfigure}[b]{0.49\textwidth} 	\includegraphics[scale=0.40, clip]{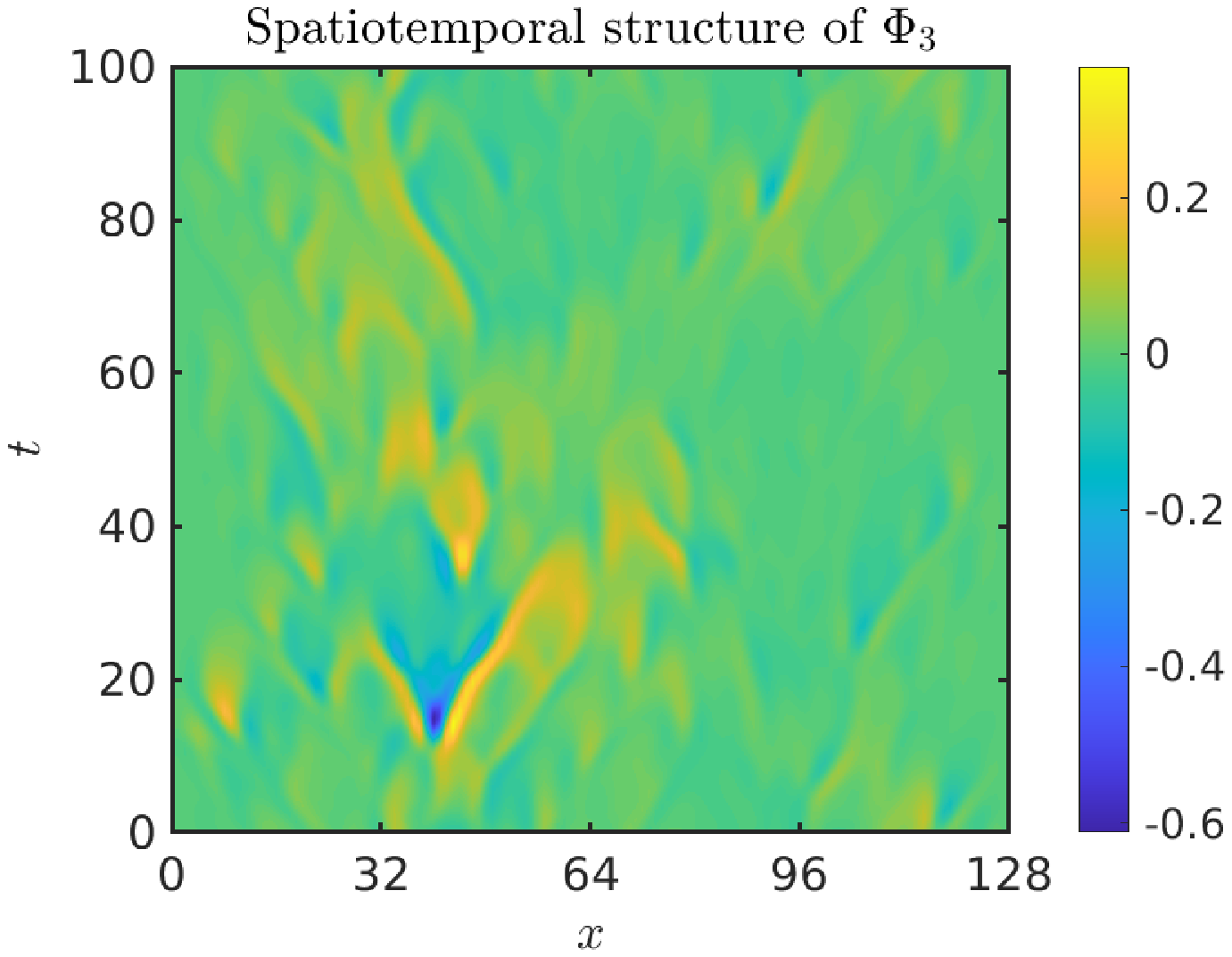}
\caption{}
\label{fig:Spatiotemporal_PHI3}
\end{subfigure}
\caption{Contour plots of (a) the forcing $\frac{df}{dc}(x,t)$, the spatio-temporal structure of optimal forcings (b)  $\bphi_1$, (c)  $\bphi_2 $ and (d)  $\bphi_3$ for $T = 100$, $c = 0$ and $f \in [-0.3,0.3]$.}
\label{fig:Spatiotemporal Singular Vectors Forcings}
\end{figure}

Contours of the actual forcing $\frac{df}{ds}(x,t)$ in the $x-t$ plane are shown in figure \ref{fig:dfds_of_t_spatiotemporal} for the same realisation as in the previous figure.  The spatio-temporal distributions of the optimal forcings that correspond to the first three singular values, i.e.\ $\bphi_1$,  $\bphi_2$ and  $\bphi_3$, are shown in panels (b)-(d) respectively. The distributions  are more difficult to interpret physically, but note the significant differences with respect to the actual forcing. It is interesting to note for example that they do not exhibit the wavy pattern of $\frac{df}{ds}(x,t)$, instead they are relatively smooth but with some local peaks and valleys. 

\begin{figure}[!htb]
\centering
\includegraphics[scale=0.5, clip]{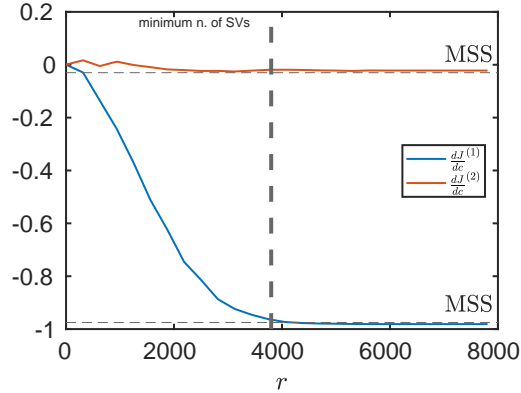}
\caption{Approximations $\left(\frac{d \overline{J_1}}{dc}\right)_{(r)}$ and $\left(\frac{d \overline{J_2}}{dc}\right)_{(r)}$ as a function of the number of retained singular values, $r$, for $c = 0$, $q = 30$ and $T = 100$.}
\label{fig: Full SVD dJu dJu2 Convergence}
\end{figure}

Using the $r$ largest singular values of $\mathcal{H}$, an approximate solution of system \eqref{eq:harmonic_balancing_system} can be written as
\begin{equation}
\label{eq:large_system_sv_solution}
\widehat{\mathcal{V}}_{(r)} = \sum_{i=1}^r \sigma_i\langle \widehat{\mathcal{R}} \bphi_i  \rangle \bpsi_i,
\end{equation}
\noindent where $\langle \widehat{\mathcal{R}}  \bphi_i  \rangle=  \bphi_i^*  \widehat{\mathcal{R}}$ is the projection of the right hand side $\widehat{\mathcal{R}}$ onto the optimal forcing $\bphi_i$. 
For $r=(2q+1)(N_u+1)$ the approximation is exact. Using $\widehat{\mathcal{V}}_{(r)}$, approximations $\left(\frac{d \overline{J_1}}{dc}\right)_{(r)}$ and $\left(\frac{d \overline{J_2}}{dc}\right)_{(r)}$ can be computed from \eqref{eq:sensitivity_J1_freq} and \eqref{eq:sensitivity_J2_freq} respectively; these are plotted as functions of $r$ in figure \ref{fig: Full SVD dJu dJu2 Convergence}. It can be seen that a relatively large value of $r$ is required to obtain an accurate result.  Although the first few singular values $\sigma_i$ are large, the component of  $\widehat{\mathcal{R}}$ along the $\bphi_i$ direction is weak, thus a substantial number of terms are required to obtain the correct sensitivity. Bearing in mind that $\widehat{\mathcal{R}}_k = \left [ \hat{\frac{d \bf_k}{d \bs}}, 0 \right ]^\top$, this is related to the different patterns between $\frac{df}{ds}(x,t)$ and optimal forcings (at least for the first 3 modes) as shown in the previous figure \ref{fig:Spatiotemporal Singular Vectors Forcings}. 
 
The linear system \eqref{eq:harmonic_balancing_system} has large storage requirements and is time consuming to solve. Below we propose an approach that can mitigate these  requirements.  

\section{A resolvent-based iterative method for the shadowing direction}\label{Simplified Fourier}
Instead of solving directly system \eqref{eq:harmonic_balancing_system}, an iterative method can be devised, where only a few diagonals are retained in the left hand side and treated implicitly, and the rest are moved to the right hand side and  updated at every iteration. For example, by retaining only the blocks of the main diagonal, the Hill matrix becomes diagonal and the blocks decouple. In this case, the iterative method takes the from 
\begin{gather}
  \begin{bmatrix}
i k \omega_0  \mathcal{I}_{N_u} -\overline{\frac{\partial \bf}{\partial \bu}}  &
 - \overline{\bf} \\
   -\overline{\bf}^\top &
   0 
   \end{bmatrix}
   \begin{bmatrix}
   \hat{\bv}_k \\ \hat{\eta}_k
   \end{bmatrix}^{(m)}
   =
   \begin{bmatrix}
    \hat{\frac{d \bf_k}{d \bs}} \\ 0
   \end{bmatrix}+
    \begin{bmatrix}
    \hat{\boldsymbol{g}}_k\\ \hat{\boldsymbol{h}}_k
   \end{bmatrix}^{(m-1)},
     \label{eq:simplied_Hill}
\end{gather}
\noindent where $m$ is the iteration number, and $\hat{\boldsymbol{g}}_k$, $\hat{\boldsymbol{h}}_k$ denote the explicitly treated terms. It is instructive to derive the form of vectors ${\boldsymbol{g}}(t)$ and ${\boldsymbol{h}}(t)$ in the time domain. To this end, using Reynolds decomposition,  
\begin{equation}
\begin{aligned}
& \bv = \overline{\bv}+ \bv', \quad \eta= \overline{\eta}+\eta', \quad \bf =  \overline{\bf} + \bf', \\
& \frac{\partial \bf}{\partial \bu} =\overline{\frac{\partial \bf}{\partial \bu}} + \frac{\partial \bf'}{\partial \bu}, \quad 
 \frac{\partial \bf}{\partial \bs} =\overline{\frac{\partial \bf}{\partial \bs}} + \frac{\partial \bf'}{\partial \bs}
\end{aligned}
\label{Reynolds}
\end{equation}
and substituting in \eqref{eq:ODE_tangent_dilation} and 
\eqref{eq:orthogonality} we get
\begin{subequations}
\begin{align}
&\frac{d \bv'}{dt} = \left (\overline{\frac{\partial \bf}{\partial \bu}} + \frac{\partial \bf'}{\partial \bu} \right )  \left ( \overline{\bv}+ \bv' \right )+ \overline{\frac{\partial \bf}{\partial \bs}} + \frac{\partial \bf'}{\partial \bs} + \left(\overline{\eta}+\eta'\right)\left(\overline{\bf}+ \bf'\right) \\
&  \langle  \left(\overline{\bf} + \bf'\right)\left(\overline{\bv}+ \bv'\right) \rangle=
\left( {\overline{\bf}}^\top + {\bf'}^\top \right) \left(\overline{\bv}+ \bv'\right) = 0 
\end{align}
\label{ODE_tangent_Reynolds1}
\end{subequations}
Taking the time-average we obtain
\begin{subequations}
\label{ODE_tangent_TA}
\begin{align}
&\overline{\frac{\partial \bf}{\partial \bu}} \overline{\bv} + \overline{\frac{\partial \bf'}{\partial \bu} \bv'} + \overline{\frac{\partial \bf}{\partial \bs}} + \overline{ \eta} \overline{\bf}+\overline{\eta' \bf'} =  0, \\
& {\overline{\bf}}^\top \overline{\bv} + \overline{{\bf'}^\top \bv' }= 0 
\end{align}
\end{subequations}
and subtracting the two sets, we get 
\begin{subequations}
\label{ODE_tangent_fluctuations}
\begin{align}
&\frac{d \bv'}{dt} = \overline{\frac{\partial \bf}{\partial \bu}} \bv' + \frac{\partial \bf'}{\partial \bu} \overline{\bv} +
\left(\frac{\partial \bf'}{\partial \bu} \bv' - \overline{\frac{\partial \bf'}{\partial \bu} \bv'} \right) + \frac{\partial \bf'}{\partial \bs} + \overline{\eta} \bf' +\eta' \overline{\bf}+ \left(\eta' \bf'- \overline{\eta' \bf'} \right) \\
& {\overline{\bf}}^\top \bv' + {\bf'}^\top \overline{\bv}+ \left({\bf'}^\top \bv' - \overline{{\bf'}^\top \bv' } \right)= 0 
\end{align}
\end{subequations}
After some rearrangement, 
\begin{subequations}
\label{ODE_tangent_fluctuations_2}
\begin{align}
&\frac{d \bv'}{dt} - \overline{\frac{\partial \bf}{\partial \bu}} \bv' - \eta' \overline{\bf} = \frac{\partial \bf'}{\partial \bs} +
\underbrace{\frac{\partial \bf'}{\partial \bu} \overline{\bv} +
\left(\frac{\partial \bf'}{\partial \bu} \bv' - \overline{\frac{\partial \bf'}{\partial \bu} \bv'} \right) + \overline{\eta} \bf' + \left(\eta' \bf'- \overline{\eta' \bf'} \right)}_{=\boldsymbol{g'}(t)} \label{ODE_tangent_fluctuations_21}\\
& {\overline{\bf}}^\top \bv' =
\underbrace{- {\bf'}^\top \overline{\bv}- \left({\bf'}^\top \bv' - \overline{{\bf'}^\top \bv' } \right)}_{=\boldsymbol{h'}(t)} \label{ODE_tangent_fluctuations_22}
\end{align}
\end{subequations}
Thus we get
\begin{subequations}
\label{ODE_tangent_fluctuations_3}
\begin{align}
&\frac{d \bv'}{dt} - \overline{\frac{\partial \bf}{\partial \bu}} \bv' - \eta' \overline{\bf} = \frac{\partial \bf'}{\partial \bs} +\boldsymbol{g'}(t) \\
& {\overline{\bf}}^\top \bv' =\boldsymbol{h'}(t)
\end{align}
\end{subequations}
and taking the Fourier transform leads to \eqref{eq:simplied_Hill} for $k \ne 0$. Similarly, the  time-average system \eqref{ODE_tangent_TA} can be written as
\begin{subequations}
\label{ODE_tangent_TA_2}
\begin{align}
&-\overline{\frac{\partial \bf}{\partial \bu}} \overline{\bv} - \overline{ \eta} \overline{\bf} = \overline{\frac{\partial \bf}{\partial \bs}}+ \underbrace{ \overline{\frac{\partial \bf'}{\partial \bu} \bv'} +\overline{\eta' \bf'}}_{=\overline{\boldsymbol{g}(t)}} \\
& {\overline{\bf}}^\top \overline{\bv} = \underbrace{- \overline{{\bf'}^\top \bv' }}_{=\overline{\boldsymbol{h}(t)}},
\end{align}
\end{subequations}
from which we obtain the form
\begin{subequations}
\label{ODE_tangent_TA_3}
\begin{align}
&-\overline{\frac{\partial \bf}{\partial \bu}} \overline{\bv} - \overline{ \eta} \overline{\bf} = \overline{\frac{\partial \bf}{\partial \bs}}+\overline{\boldsymbol{g}(t)}, \\
& {\overline{\bf}}^\top \overline{\bv} =\overline{\boldsymbol{h}(t)},
\end{align}
\end{subequations}
that corresponds to $k=0$ for system \eqref{eq:simplied_Hill}. 

As can be seen from \eqref{ODE_tangent_fluctuations_21}, ${\boldsymbol{g'}(t)}$ consists of two groups of three terms; the first group involves the fluctuating Jacobian and sensitivity $\bv'$, and the second the fluctuating $\eta'$ and $\bf'$. It is possible to get a simplified system by assuming that $\eta$ is constant (for the physical interpretation see \cite{LASAGNA2019119}). In this case we get
\begin{subequations}
\label{ODE_periodic_shadowing_1}
\begin{align}
&\frac{d \bv'}{dt} - \overline{\frac{\partial \bf}{\partial \bu}} \bv' = \frac{\partial \bf'}{\partial \bs} +  \eta \bf'+
\underbrace{\frac{\partial \bf'}{\partial \bu} \overline{\bv} +
\left(\frac{\partial \bf'}{\partial \bu} \bv' - \overline{\frac{\partial \bf'}{\partial \bu} \bv'} \right)}_{=\boldsymbol{\tilde{g}'}(t)} \label{ODE_periodic_shadowing_1a}\\
&-\overline{\frac{\partial \bf}{\partial \bu}} \overline{\bv} - \eta \overline{\bf} = \overline{\frac{\partial \bf}{\partial \bs}}+ \underbrace{ \overline{\frac{\partial \bf'}{\partial \bu} \bv'} }_{=\overline{\boldsymbol{\tilde{g}}(t)}}
\label{ODE_periodic_shadowing_1b}
\end{align}
\end{subequations}
This system requires an additional constraint to obtain the constant $\eta$. To this end, we require that $\bf(t)$ and $\bv(t)$ are perpendicular in a time-average sense, i.e. 
\begin{equation}
\frac{1}{T}\int_0^T\bf^\top(t)  \bv(t) dt=
\sum _{k=-q}^{q} \hat{\bf}_{-k}^\top \hat{\bv}_k = \sum _{k=-q}^{q}  \hat{\bf}_k^* \hat{\bv}_k=0,
\end{equation}
because $\bf(t)$ is a real variable. Note that this condition couples together all Fourier components. Taking the Fourier transform of \eqref{ODE_periodic_shadowing_1} we can form the following iterative method
\begin{subequations}
\label{ODE_simplified_fourier_2}
\begin{align}
&\left ( i k \omega_0  \mathcal{I}_{N_u} - \overline{\frac{\partial \bf}{\partial \bu}} \right ) \hat{\bv}_{k}^{(m)} = \frac{\partial \hat{\bf}_k}{\partial \bs} + \eta^{(m)} \hat{\bf}_k + \boldsymbol{\tilde{g}}_k^{(m-1)} \quad [k=-q \dots q]
\label{ODE_simplified_fourier_21}\\
&\sum _{k=-q}^{q}  \hat{\bf}_k^* \hat{\bv}_k^{(m)}=0
\end{align}
\end{subequations}
In the first iteration $m=1$, we set  $\boldsymbol{\tilde{g}}_k^{(0)}=0$ and we get
\begin{subequations}
\label{ODE_simplified_fourier_3}
\begin{align}
&\left ( i k \omega_0  \mathcal{I}_{N_u} - \overline{\frac{\partial \bf}{\partial \bu}} \right ) \hat{\bv}_{k}^{(1)} = \frac{\partial \hat{\bf}_k}{\partial \bs} + \eta^{(1)} \hat{\bf}_k \quad [k=-q \dots q] 
\label{eq:ODE_simplified_fourier_first_iteration}\\
&\sum _{k=-q}^{q}  \hat{\bf}_k^* \hat{\bv}_k^{(1)}=0
\label{eq:ODE_simplified_fourier_first_iteration_2}
\end{align}
\end{subequations}
These two equations can be combined together to obtain $\eta^{(1)}$. Denoting the standard resolvent operator as 
\begin{equation}
\mathcal{R}(k\omega_0)=\left ( i k \omega_0  \mathcal{I}_{N_u} - \overline{\frac{\partial \bf}{\partial \bu}} \right )^{-1},
\end{equation}
solving for $\hat{\bv}_{k}^{(1)}$ and substituting in \eqref{eq:ODE_simplified_fourier_first_iteration_2} we get
\begin{equation}
\sum _{k=-q}^{q}  \hat{\bf}_k^* \left[
\mathcal{R}(k\omega_0) \left( \frac{\partial \hat{\bf}_k}{\partial \bs} + \eta^{(1)} \hat{\bf}_k \right) \right]=0
\end{equation}
from which we obtain
\begin{equation}
\eta^{(1)}=-\frac{\sum _{k=-q}^{q}  \hat{\bf}_k^* 
\mathcal{R}(k\omega_0) \frac{\partial \hat{\bf}_k}{\partial \bs} }
{\sum _{k=-q}^{q}  \hat{\bf}_k^* 
\mathcal{R}(k\omega_0) \hat{\bf}_k}
\end{equation}
or
\begin{equation}
\eta^{(1)}=-\frac{\sum _{k=-q}^{q}  \hat{\bf}_k^* \boldsymbol{\lambda}_k}
{\sum _{k=-q}^{q}  \hat{\bf}_k^* 
\boldsymbol{\mu}_k}
\end{equation}
where
\begin{subequations}
\label{eq:lambda_k_and_mu_k}
\begin{align}
&\mathcal{R}(k\omega_0) \frac{\partial \hat{\bf}_k}{\partial \bs}= \boldsymbol{\lambda}_k 
\Rightarrow
\left ( i k \omega_0  \mathcal{I}_{N_u} - \overline{\frac{\partial \bf}{\partial \bu}} \right ) \boldsymbol{\lambda}_k=  \frac{\partial \hat{\bf}_k}{\partial \bs}
\label{eq:lambda_k}\\
& \mathcal{R}(k\omega_0) \hat{\bf}_k=\boldsymbol{\mu}_k
\Rightarrow
\left ( i k \omega_0  \mathcal{I}_{N_u} - \overline{\frac{\partial \bf}{\partial \bu}} \right ) \boldsymbol{\mu}_k=   \hat{\bf}_k
\label{eq:mu_k}
\end{align}
\end{subequations}
Thus the solution of two linear systems that involve the standard resolvent operator, $\mathcal{R}(k\omega_0)$, is required. Due to the linearity of \eqref{ODE_simplified_fourier_21}, 
\begin{equation}
\hat{\bv}_{k}^{(1)}=\boldsymbol{\lambda}_k+\eta^{(1)} \boldsymbol{\mu}_k.
\label{eq:decomposition_of_vk}
\end{equation}

Applying inverse Fourier transform  to $\hat{\bv}_{k}^{(1)}$ yields ${v}^{(1)}(t)$, from which ${\boldsymbol{\tilde{g}}(t)}$ can be obtained from \eqref{ODE_periodic_shadowing_1}, and Fourier transformed to find  $\boldsymbol{\tilde{g}}_k^{(1)}$. The right hand side of \eqref{ODE_simplified_fourier_21} can then be assembled and the second iteration  performed. Note that system \eqref{eq:mu_k} does change with  $m$, thus $\boldsymbol{\mu}_k$ is computed once. In this process, the key variables $\boldsymbol{\lambda}_k$, $\eta$,  and $\boldsymbol{\mu}_k$ required for the evaluation of the sensitivity $\hat{\bv}_{k}$ (also known as shadowing direction) are obtained with the aid of the Resolvent operator $\mathcal{R}(k\omega_0)$. This is therefore a resolvent-based iterative method for computing the shadowing direction, called Resolvent-based Shadowing (RbS).

The most efficient approach is to perform LU decomposition of $\mathcal{R}(k\omega_0)$ once at the start of the sensitivity analysis and solve systems \eqref{eq:lambda_k_and_mu_k} with forward and backward substitution for each $k$.
Note that the decomposition \eqref{eq:decomposition_of_vk} of $\hat{\bv}_{k}$ into a linear combination  of $\boldsymbol{\lambda}_k$ and $\boldsymbol{\mu}_k$ is valid for all iterations, and of course the final converged solution.
 
In \eqref{ODE_periodic_shadowing_1a}, the term  $\frac{\partial \bf'}{\partial \bu} \overline{\bv}$ was considered as part of $\boldsymbol{\tilde{g}'}(t)$ and treated explicitly. This is not necessary, but it simplifies the algebra. A better approach would be to obtain $\overline{\bv}$ from \eqref{ODE_periodic_shadowing_1b} and substitute in \eqref{ODE_periodic_shadowing_1a}; this would lead to a form very similar to \eqref{eq:ODE_simplified_fourier_first_iteration}, albeit more complex. The process to extract $\eta$  remains the same. Such an approach would couple better the time-average and the fluctuating components of the solution with an expected  beneficial effect on the convergence rate.  

The storage requirements of the proposed iterative method are much reduced compared with the harmonic balancing approach presented in section  \ref{sensitivity analysis}. Only the time-average Jacobian $\overline{\frac{\partial \bf}{\partial \bu}}$ needs to be stored in sparse matrix form in order to assemble the standard resolvent operator $\mathcal{R}(k\omega_0)$. The computational bottleneck is the application of LU decomposition to $\mathcal{R}(k\omega_0)$, which needs however to be performed only once. If the method is applied to  fluid flow problems that possess at least one direction of homogeneity (for example boundary layers, jets, pipe flows, flows around bluff bodies etc), this decomposition can be performed very efficiently using existing linear algebra packages, such as MUMPS \cite{MUMPS}, that exploit the sparse structure of the Jacobian. For three-dimensional inhomogeneous flows this is more challenging, but doable (at least for moderate-size systems).  

\begin{figure}[!htb]
\centering
\begin{subfigure}[b]{0.49\textwidth} 	\includegraphics[scale=0.40, clip]{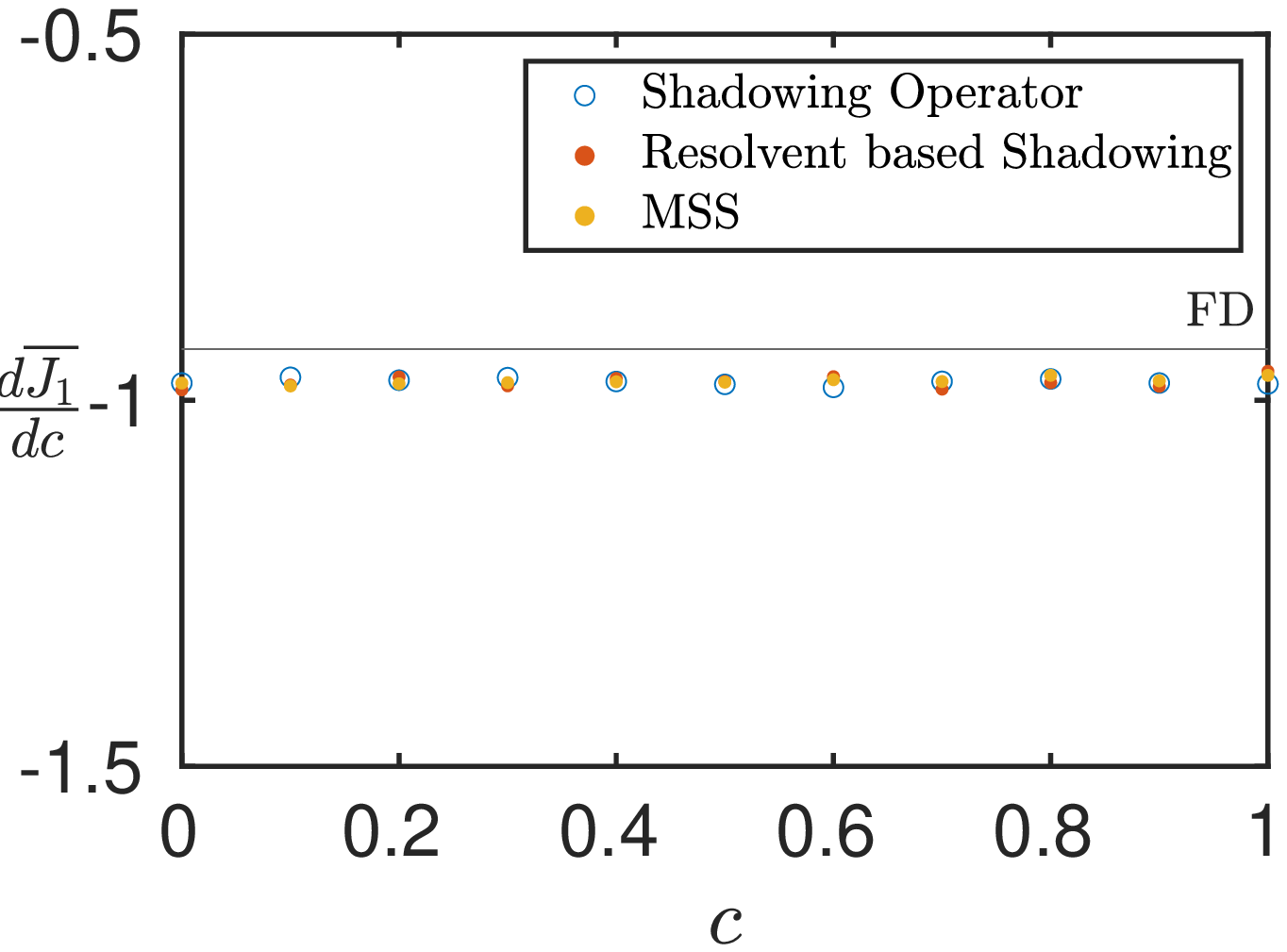}
\caption{}
\label{fig:top-left}
\end{subfigure}
\begin{subfigure}[b]{0.49\textwidth} 	\includegraphics[scale=0.40, clip]{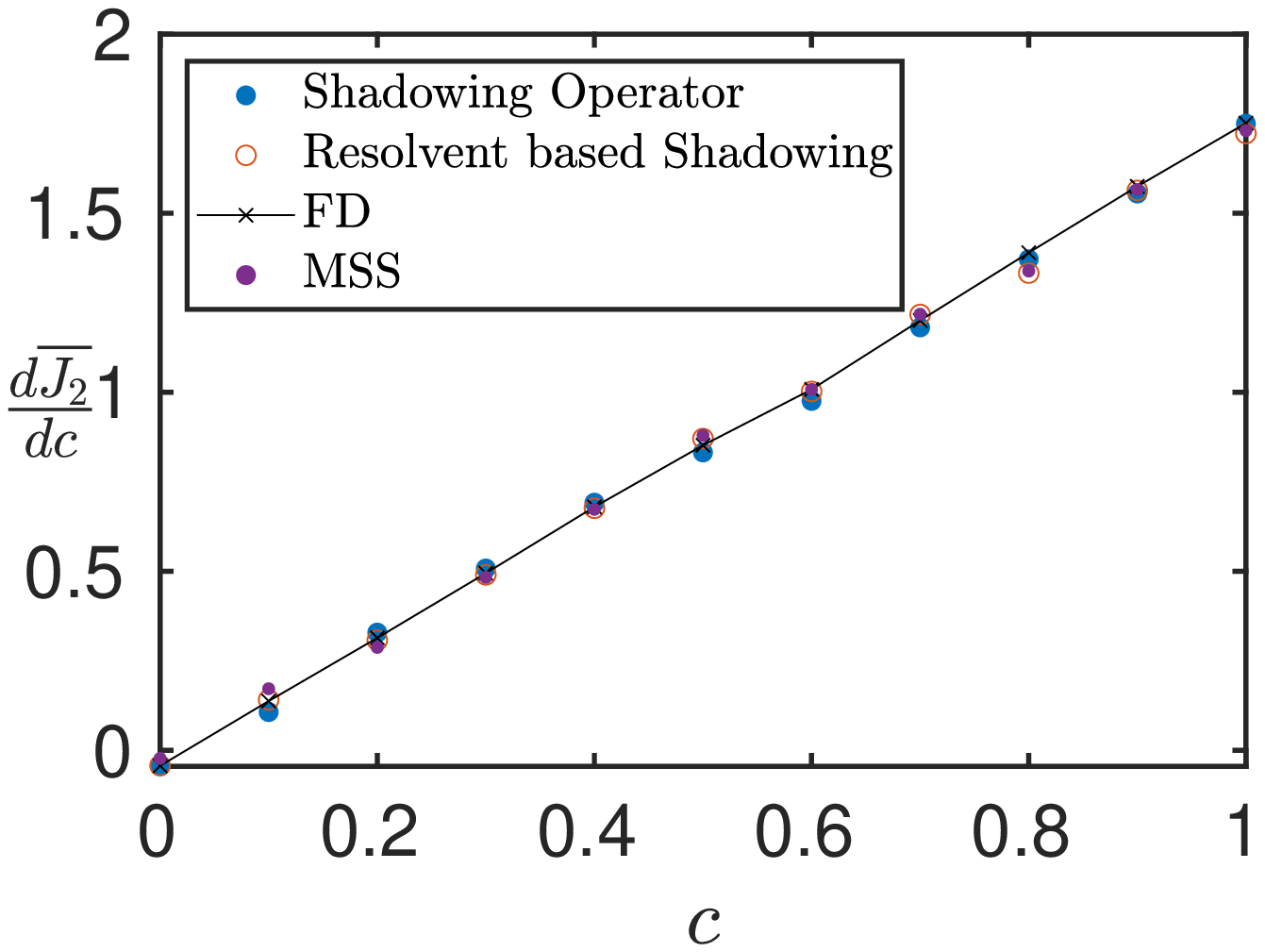}
\caption{}
\label{fig:top-right}
\end{subfigure}
\caption{Comparison of sensitivities obtained using full harmonic balancing, finite differences, MSS and Resolvent based Shadowing (RbS) for (a) $\frac{d \overline{J_1}}{dc}$, (b) $\frac{d \overline{J_2}}{dc}$. Results are averaged over 100 random initial conditions and the time-average Jacobian was obtained using $T=500$.}
\label{KS_FD_comparison}
\end{figure}

The proposed method was applied to compute the sensitivities of the objective functions  $\overline{J_1}(c)$ and $\overline{J_2}(c)$, defined  in \eqref{KS objective u} and \eqref{KS objective 2}, for the Kuramoto-Sivashinsky equation. Since $\eta$ is constant, equations \eqref{eq:sensitivity_J1_freq} and \eqref{eq:sensitivity_J2_freq} are simplified to
\begin{equation}
\frac{d \overline{J_1}}{dc} = \frac{1}{L} \int_0^L  \hat{\bv}_0 \, dx,
\label{eq:sensitivity_J1_freq_simplified}
\end{equation}
and
\begin{equation}
\frac{d \overline{J_2}}{dc} = \frac{2}{L} \int_0^L   \sum_{k=-q}^{q} \hat{\bu}_{-k} \hat{\bv}_k \, dx=
 \sum_{k=-q}^{q} \frac{2}{L} \int_0^L   \hat{\bu}^*_k \hat{\bv}_k \, dx,
\label{eq:sensitivity_J2_freq_simplified}
\end{equation}
respectively. A comparison of the sensitivities produced by the  harmonic balance approach of section \ref{sensitivity analysis}, finite differences, and the methodology presented in this section (applied with a single iteration) are shown in figure \ref{KS_FD_comparison}. 
The sensitivities match well with the those obtained with shadowing harmonic for both objective functions. The explanation for this (perhaps surprisingly) good result for  $\frac{d \overline{J_1}}{dc}$ is provided in figure \ref{fig: Comparison Full Decoupled vTA}, where we compare the distribution of  $\hat{\bv}_0(x)=\overline{\bv}(x)$ between the two approaches. As can be seen, after a single iteration, the average value over $x$ is approximated well, but the details of the distribution in the middle of the domain are not captured. However, these differences cancel out when integrating  $\hat{\bv}_0(x)$, see \eqref{eq:sensitivity_J1_freq_simplified}, leading to an accurate sensitivity value. If $\overline{J_1}(c)$ had been formulated as the average of $\overline{u}(x;c)$ over a smaller part of the domain, then the sensitivities with a single iteration would not have been accurate. Of course,  in this case additional iterations can be performed. 

In terms of computational cost, it took approximately $0.15$s of CPU time to evaluate these sensitivities; this is about two orders of magnitude faster compared to preconditioned MSS and one order faster with respect to the shadowing harmonic approach. Again, we stress that these results are case dependent. More research is need to investigate the performance of the algorithm in other flow cases.

In the next section, we explore the properties of the standard resolvent operator $\mathcal{R}(k\omega_0)$.

\section{Analysis of the sensitivities using of the standard resolvent operator}
\label{Resolvent Decoupled}
\begin{figure}[!htb]
\centering
\includegraphics[scale=0.70, clip]{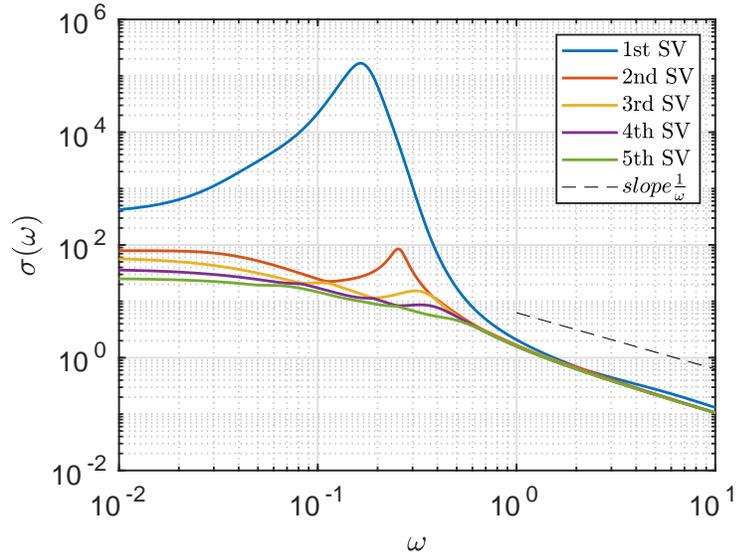}
\caption{Variation of the five largest singular values against $\omega$, for $c = 0$, $T = 100$.}
\label{fig:KS_singular_values}
\end{figure}

The five largest singular values of the resolvent operator $\mathcal{R}(\omega)$ are plotted in figure \ref{fig:KS_singular_values} as a function of $\omega$. The maximum singular value has a peak at $\omega |_{\sigma = \sigma_{max} } \approx 0.164$ and in this frequency region it is more than 3 orders of magnitude larger than the second. This result is compatible with the spectra of figure \ref{fig: FFT Spectra}; large gains appear in the frequency range with strong spectral content. For angular frequencies $\omega > 2\, \pi \,0.3 \approx 1.9$ there is no coupling with the time-average Jacobian, leading to $\sigma(\omega) \approx 1/\omega$ (indicated by a black dashed line); again this is consistent with the frequency spectra.

\begin{figure}[!htb]
\centering
\begin{subfigure}[b]{0.49\textwidth} 	\includegraphics[scale=0.40, clip]{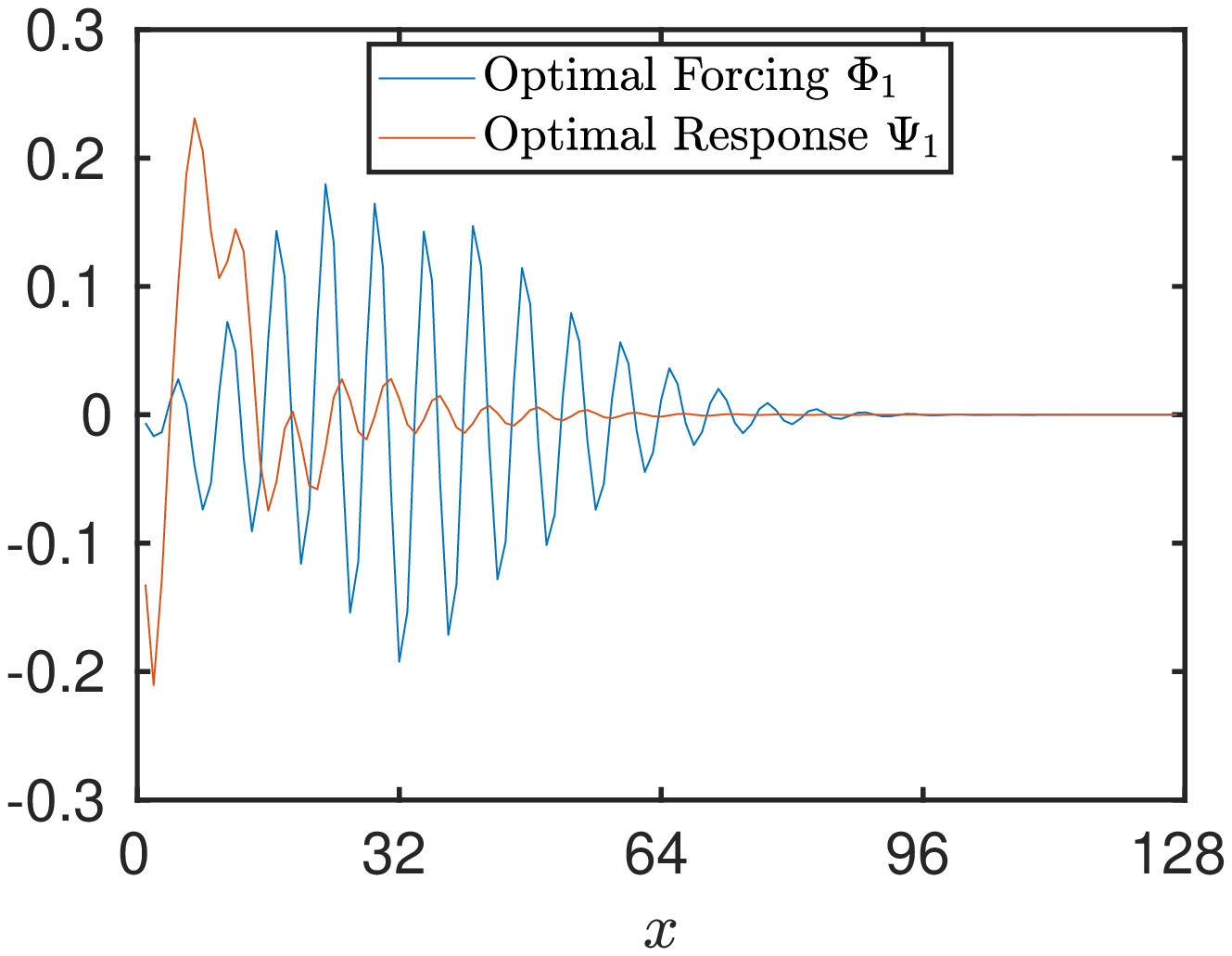}
\caption{}
\label{fig: SVD Optimal Forcing f1}
\end{subfigure}
\begin{subfigure}[b]{0.49\textwidth} 	\includegraphics[scale=0.40, clip]{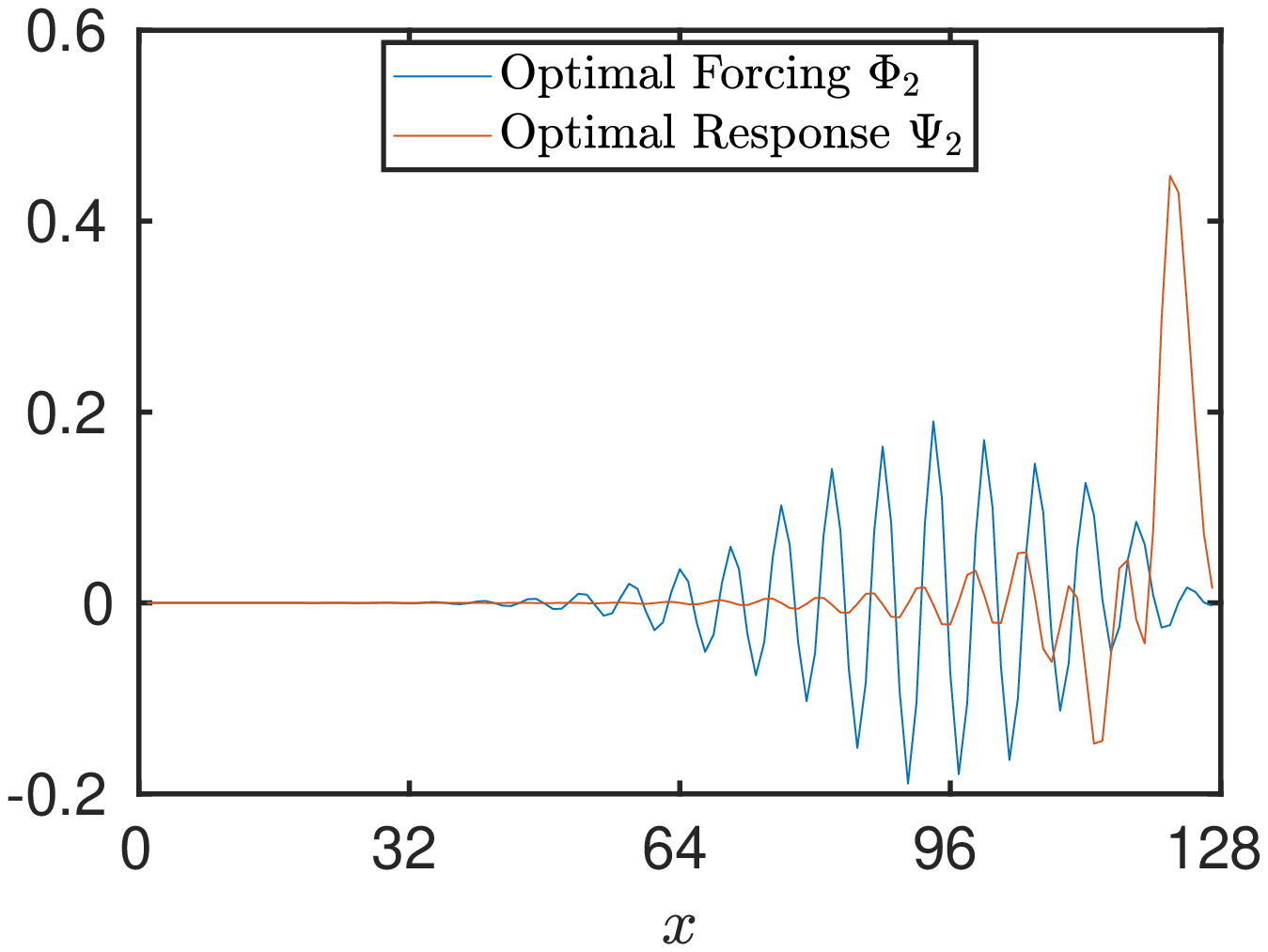}
\caption{}
\label{fig: Optimal Forcing f2}
\end{subfigure}
\caption{Optimal forcing and response corresponding to the largest (left) and second largest (right) singular values for $ \omega=0.164$, $T = 20000$ and $c = 0$.}
\label{fig: SVD Optimal Forcing }
\end{figure}

In figure \ref{fig: SVD Optimal Forcing } the optimal forcing and response that correspond to the two largest singular values for $ \omega=0.164$ are plotted. These plots are easier to interpret compared to the spatio-temporal maps shown in figures \ref{fig:Spatiotemporal Singular Vectors Responses} and \ref{fig:Spatiotemporal Singular Vectors Forcings}. For example, note that both distributions are spatially localised, on the left half of the domain for the largest singular value, and on the right half for the second largest. In both cases, the optimal forcing is located upstream of the optimal response, as expected due to the convective nature of the KS (recall that $\overline{u}(x)$, shown in figure \ref{KS_time_averages}, changes sign in the middle of the domain). 

Substituting expression  \eqref{eq:decomposition_of_vk} into \eqref{eq:sensitivity_J1_freq_simplified} we get 
\begin{equation}
\frac{d \overline{J_1}}{dc} = \frac{1}{L} \int_0^L \Big ( \boldsymbol{\lambda}_0 + \eta \boldsymbol{\mu}_0 \Big ) dx.
\label{Res: Sensitivity 2}
\end{equation}

The solutions $ \lambda_0$ and $\mu_0$ of the linear systems \eqref{eq:lambda_k} and \eqref{eq:mu_k} can be written in terms of the optimal forcings and responses evaluated for $\omega=0$ as  
\begin{equation}
{\lambda_0}_{(r)}=\sum_{i=1}^{r} \sigma_i^{(\omega = 0)}\left< \overline{\frac{\partial \bf}{\partial \bs}} \Phi^{(\omega = 0)}_i \right> \Psi_i^{(\omega = 0)}, \quad {\mu_0}_{(r)}=\sum_{i=1}^{r} \sigma_i^{(\omega = 0)}\left< \overline{\bf} \Phi_i^{(\omega = 0)}  \right> \Psi_i^{(\omega = 0)},
\label{eq:approx_lambda0_mu0}
\end{equation}
where $r$ is the number of retained singular values. This expression is analogous to equation \eqref {eq:large_system_sv_solution} presented earlier for the harmonic balance method.  Substituting in \eqref{Res: Sensitivity 2} we get
\begin{equation}
\left(\frac{d \overline{J_1}}{dc}\right)_{(r)}
= \sum_{i=1}^{r} \sigma_i^{(\omega = 0)} 
\left ( \left< \overline{\frac{\partial \bf}{\partial c}} \Phi_i^{(\omega = 0)} \right> + 
\eta 
\left< \overline{\bf} \Phi_i^{(\omega = 0)} \right>
\right ) 
\frac{1}{L} \int_0^L 
 \Psi_i dx.
\label{Res: Sensitivity 3}
\end{equation}
The sensitivity can therefore be written as the weighted sum of the spatial averages of the optimal responses, $\Psi_i$.

\begin{figure}[!htb]
\centering
\begin{subfigure}[b]{0.49\textwidth} 	\includegraphics[scale=0.43, clip]{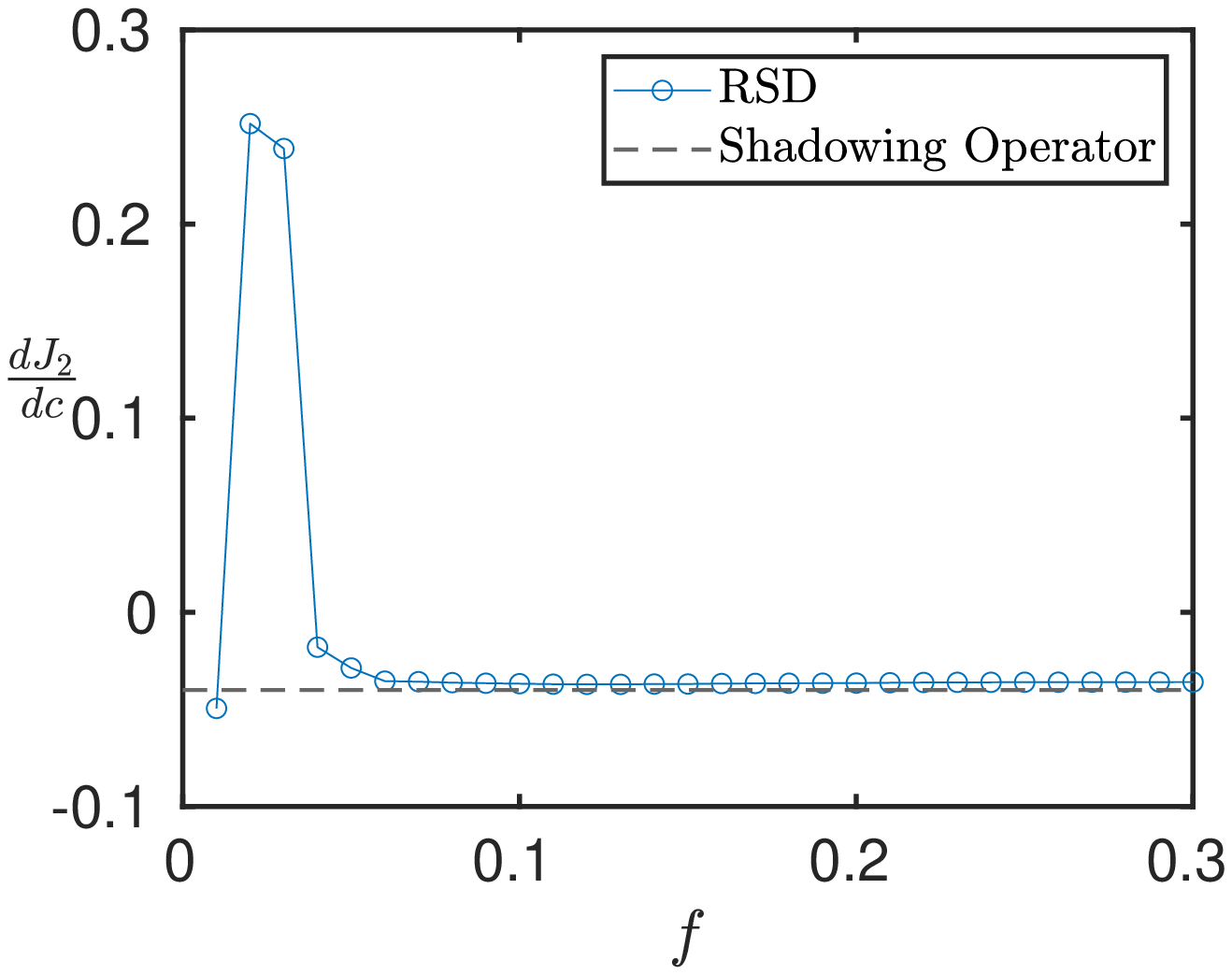}
\caption{}
\label{fig:KS_spectral_convergence_fourier}
\end{subfigure}
\begin{subfigure}[b]{0.49\textwidth} 	\includegraphics[scale=0.43, clip]{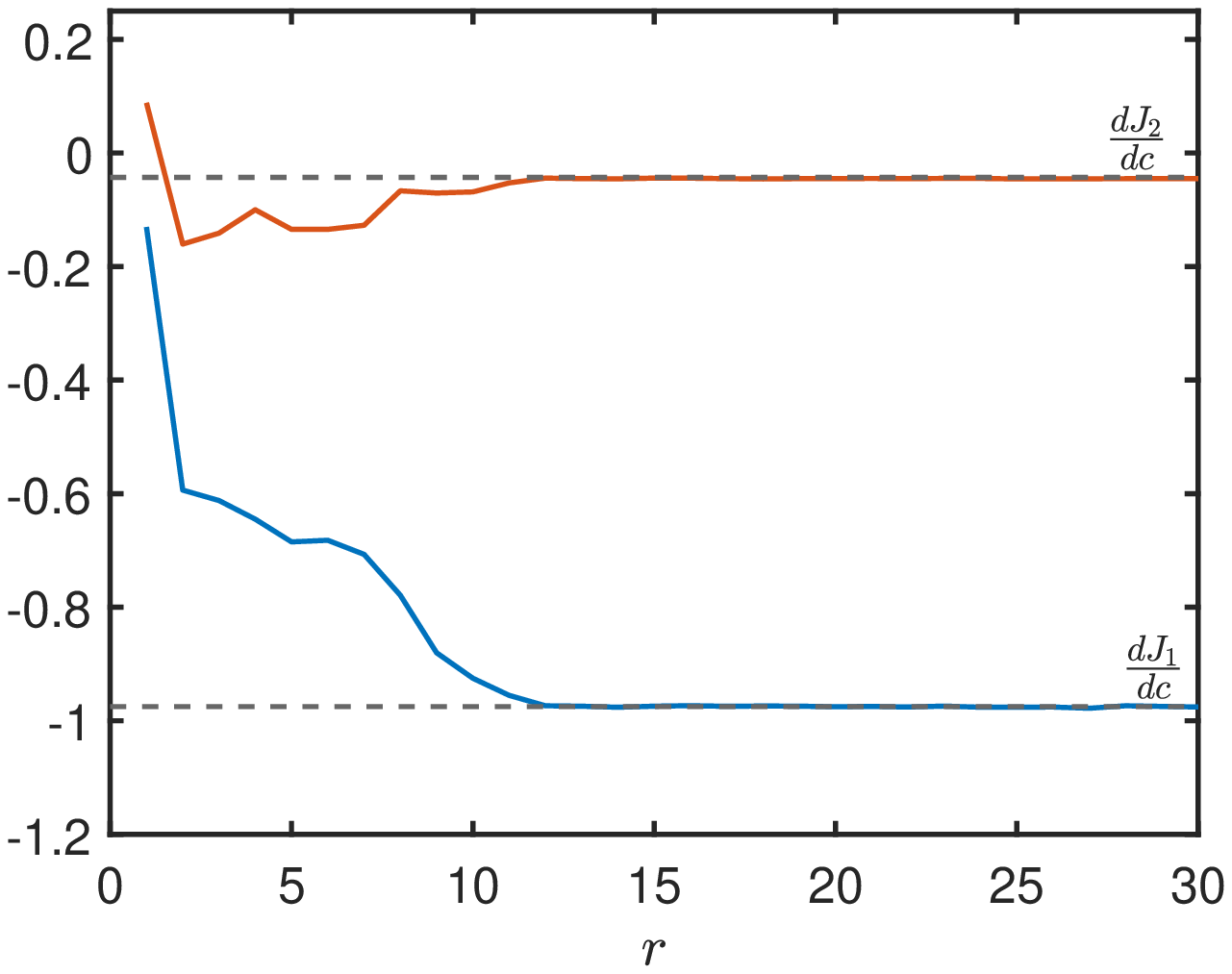}
\caption{}
\label{fig:KS_SV_congergence_Ju2}
\end{subfigure}
\caption{(a) Convergence of $\frac{d\overline{J_2}}{dc}$ computed using frequencies in the interval $[-f,f]$ as $f$ increases, for $c = 0$ and $T = 100$. (b) Effect of the number of retained singular values, $r$, on the convergence of sensitivities \eqref{KS objective u} and  eq. \eqref{KS objective 2} for $T = 100$, $q = 30$ and $c = 0$. }
\label{fig:KS_SV_congergence}
\end{figure}

Similarly for $\frac{d \overline{J_2}}{dc}$, we get from \eqref{eq:sensitivity_J2_freq_simplified}, 
\begin{equation}
\left(\frac{d \overline{J_2}}{dc}\right)_{(r)} =\sum_{k=-q}^{q} 
\frac{2}{L} \int_0^L   \bu_{k}^* \left ( \lambda_k  + \eta \mu_k \right )  dx,
\label{Res: Sensitivity 4}
\end{equation}
where
\begin{subequations}
\begin{align}
&{\lambda_k}_{(r)}=\sum_{i=1}^{r} \sigma_i^{(\omega = k \omega_0)}\left< \frac{\partial \hat{\bf}_k}{\partial c} \Phi_i^{(\omega = k \omega_0)} \right> \Psi_i^{(\omega =k \omega_0)}, \\ &{\mu_0}_{(r)}=\sum_{i=1}^{r} \sigma_i^{(\omega = k \omega_0)} \left< \hat{\bf}_k \Phi_i^{(\omega = k \omega_0)}  \right> \Psi_i^{(\omega =k \omega_0)}.
\end{align}
\label{eq:approx_lambdak_muk}
\end{subequations}
Expression \eqref{Res: Sensitivity 4} is useful because it allow us to find the contribution of each frequency on the sensitivity; we investigate this in figure \ref{fig:KS_spectral_convergence_fourier}. More specifically, we compute $\frac{d\overline{J_2}}{dc}$ using frequencies in the interval $[-f,f]$, where $f=\omega/(2\pi)$, and we plot the result against $f$. Note the convergence of $\frac{d\overline{J_2}}{dc}$ to the value predicted by the harmonic resolvent as $f$ increases, i.e.\ the range of $k$ in \eqref{Res: Sensitivity 4} expands. It can be seen that only the frequency range $[0.01,0.07]$ contributes, which is consistent with the spectra shown in figure \ref{fig: FFT Spectra}.  Outside this range, the spectral content is small, and does not contribute to the sensitivity. Note that this is the result of a single realisation; no averaging over initial conditions has been performed to obtain this plot. 

\begin{figure}[!htb]
	\centering
	\includegraphics[scale=0.45, clip]{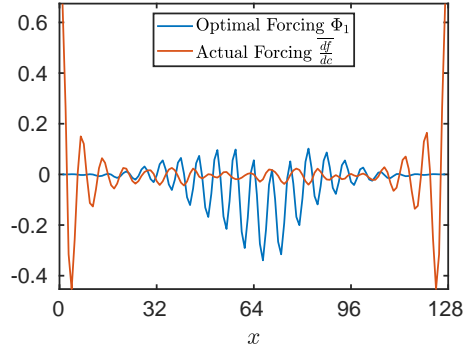}
	\caption{Optimal forcing $\Phi_1$ for $ \omega= 0$ and time-average forcing $\overline{\frac{\partial \bf}{\partial c}}$ against $x$, for $T = 20000$, $c=0$. }
	\label{Optimal_Forcing_vs_actual_forcing}
\end{figure}

In figure \ref{fig:KS_SV_congergence_Ju2} we plot the convergence of the two sensitivities against the number of retained singular values $r$ (same $r$ for every $k$) in the summations \eqref{Res: Sensitivity 3} and \eqref{eq:approx_lambdak_muk}. Approximately 10-12 singular values are required for convergence. Hence, although the maximum singular value is significantly larger compared with the rest as evidenced in \ref{fig:KS_singular_values}, keeping just one contribution will not provide accurate results. In order to explain this behaviour for $\frac{d\overline{J_1}}{dc}$, the optimal forcing corresponding to $\sigma_{max}$ at $\omega=0$ is plotted together with the true forcing $\overline{\frac{\partial \bf}{\partial c}}$ in figure \ref{Optimal_Forcing_vs_actual_forcing}. The latter has strong footprint in the boundaries of the domain, while the former in the centre, thus the projection $\left< \overline{\frac{\partial \bf}{\partial c}} \Phi^{(\omega = 0)}_1 \right>$ appearing in expression \eqref{eq:approx_lambda0_mu0} for ${\lambda_0}_{(r)}$ is weak and mitigates the effect of the large $\sigma_{max}$.  

\begin{figure}[!htb]
\centering
\begin{subfigure}[b]{0.49\textwidth} 	\includegraphics[scale=0.40, clip]{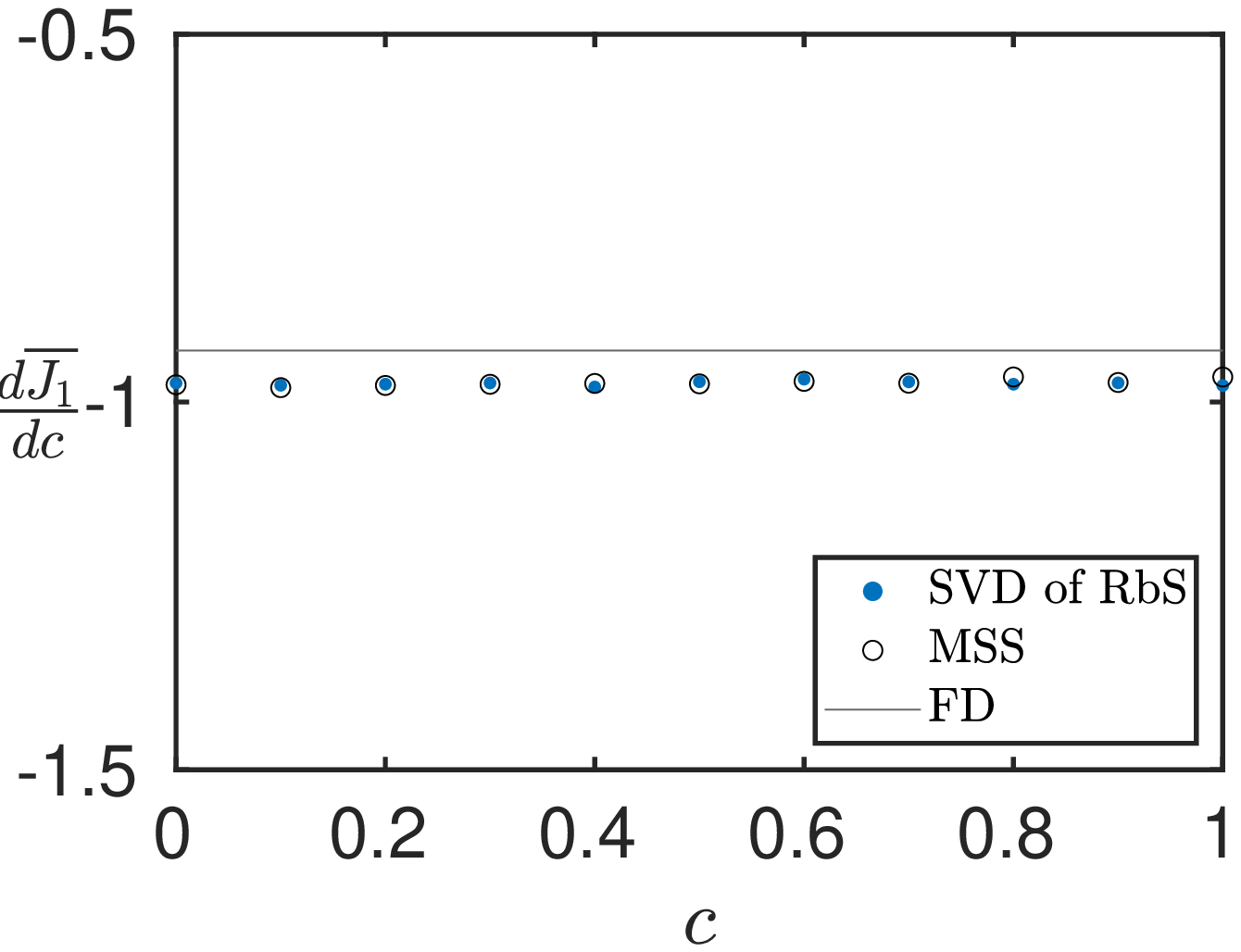}
\caption{}
\label{fig:KS_SVD_FD_comparison_Ju}
\end{subfigure}
\begin{subfigure}[b]{0.49\textwidth} 	\includegraphics[scale=0.40, clip]{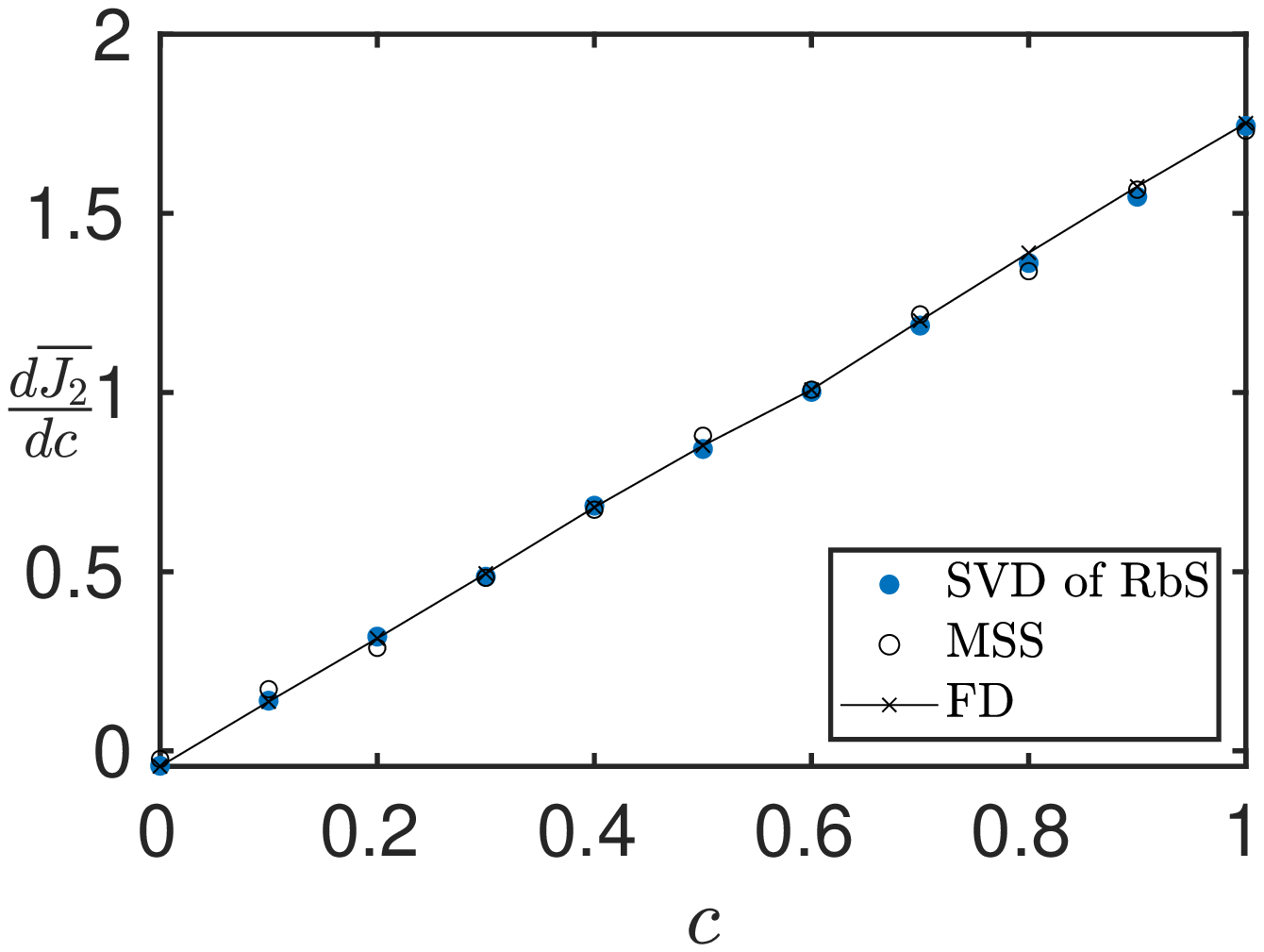}
\caption{}
\label{fig:KS_SVD_FD_comparison_Ju2}
\end{subfigure}
\caption{Comparison of the sensitivities computed with SVD of RbS, FD and MSS for (a) eq. \eqref{KS objective u} and (b) eq. \eqref{KS objective 2}, for $T = 100$,  $q = 7$ and $r = 12$. Results are averaged over 100 random initial conditions.}
\label{fig:KS_SVD_FD_comparison}
\end{figure}

In figure \ref{fig:KS_SVD_FD_comparison} we compare the sensitivities computed with MSS, RbS and finite differences for different values of $c$, using $T = 100$, $q = 7$ and $r = 12$. This combination of $T$ and $q$ captures the frequency band where fluctuations have large spectral footprint, while at each frequency a number of singular values are retained based on the evidence from figure \ref{fig:KS_SV_congergence_Ju2}. Using these settings, the  matching with the reference MSS values is very good.  

\section{Conclusions}
\label{conclusions chapter}
Least-squares shadowing (LSS) is a very promising method for the computation of sensitivities of time-average quantities of chaotic dynamical systems to system parameters. The original method and its variants have been formulated in the time domain, resulting in computational costs that scale with the number of PLEs. Hence, the application of existing formulations to complex dynanical systems with a large number of PLEs, such as high Reynolds number turbulent flows, is prohibitively expensive. To circumvent this problem, in this paper, we reformulate LSS in the frequency domain using the harmonic balancing approach. This formulation leads naturally to the definition of the shadowing harmomic operator; we apply this operator and study its properties for the Kuramoto-Sivasinky equation.  

The sensitivities computed with the time- and frequency-domain formulations match. However, using a direct method to solve the shadowing harmonic linear system has large memory and cpu requirements, which rapidly increase with the number of degrees of freedom. On the other hand, the computational cost is independent of the number of positive Lyapunov exponents. To mitigate these requirements, we propose an iterative approach, where only the diagonal blocks of the shadowing harmomic operator need to be stored and inverted. These blocks correspond to the standard resolvent operator, thus this is a resolvent-based iterative method to compute the shadowing direction. At each iteration, the Fourier components of the  shadowing direction are (partially) coupled through an appropriately formulated condition of orthogonality between the solution and the system trajectory. We show that this approach provides very accurate results even with a single iteration. Furthermore we show that this is dependent on the particular objective function considered; other functions may require more iterations. 

Clearly more work is needed to assess the accuracy, cost and scalability of the proposed iterative approach. Many questions arise for example, what are the properties of the dynamical system that allow accurate results to be obtained with few iterations? How does this depend on the parameters of the system, for example  Reynolds number? Is it related to the quantity of interest and the underlying flow structures that determine this quantity? For instance, dissipation in turbulent flows is determined by the smallest scale structures; does this mean that evaluating the sensitivity of dissipation will require more iterations compared to the sensitivity of the forces acting on a body surface (that are mostly determined by large scale structures)? How sensitive is the convergence rate to the number of positive Lyapunov exponents?  Work towards answering these questions will form part of future research.


\section*{Acknowledgements}
The first author wishes to acknowledge the financial support of the President's Scholarship Award from Imperial College London. The second author is grateful for the financial support provided by EPSRC, grant No. EP/P020194/1.

\bibliography{new}

\begin{thebibliography}{10}
\expandafter\ifx\csname url\endcsname\relax
  \def\url#1{\texttt{#1}}\fi
\expandafter\ifx\csname urlprefix\endcsname\relax\def\urlprefix{URL }\fi
\expandafter\ifx\csname href\endcsname\relax
  \def\href#1#2{#2} \def\path#1{#1}\fi

\bibitem{WANG2014210}
Q.~Wang, R.~Hu, P.~Blonigan, Least squares shadowing sensitivity analysis of
  chaotic limit cycle oscillations, Journal of Computational Physics 267 (2014)
  210 -- 224.

\bibitem{Leapaper}
D.~J. Lea, M.~R. Allen, T.~W. Haine, Sensitivity analysis of the climate of a
  chaotic system, Tellus A: Dynamic Meteorology and Oceanography 52~(5) (2000)
  523--532.

\bibitem{bewley_moin_temam_2001}
T.~R. Bewley, P.~Moin, R.~Temam, \uppercase{DNS}-based predictive control of
  turbulence: an optimal benchmark for feedback algorithms, Journal of Fluid
  Mechanics 447 (2001) 179–225.

\bibitem{xiao_papadakis_2019}
D.~Xiao, G.~Papadakis, Nonlinear optimal control of transition due to a pair of
  vortical perturbations using a receding horizon approach, Journal of Fluid
  Mechanics 861 (2019) 524–555.

\bibitem{Eyink_2004}
G.~L. Eyink, T.~W.~N. Haine, D.~J. Lea, Ruelle{\textquotesingle}s linear
  response formula, ensemble adjoint schemes and \uppercase{L}{\'{e}}vy
  flights, Nonlinearity 17~(5) (2004) 1867--1889.

\bibitem{kubo_1966}
R.~Kubo, The \uppercase{F}luctuation-\uppercase{D}issipation
  \uppercase{T}heorem, Reports on Progress in Physics 29~(1) (1966) 255.

\bibitem{Thuburn2005}
J.~Thuburn, {Climate sensitivities via a Fokker–Planck adjoint approach},
  Quarterly Journal of the Royal Meteorological Society 131~(605) (2005)
  73--92.

\bibitem{CRASKE2019243}
J.~Craske, Adjoint sensitivity analysis of chaotic systems using cumulant
  truncation, Chaos, Solitons \& Fractals 119 (2019) 243 -- 254.

\bibitem{Lasagna_2018}
D.~Lasagna, Sensitivity analysis of chaotic systems using unstable periodic
  orbits, SIAM Journal on Applied Dynamical Systems 17~(1) (2018) 547--580.

\bibitem{Wang2014ConvergenceAverages}
Q.~Wang, {Convergence of the Least Squares Shadowing Method for Computing
  Derivative of Ergodic Averages}, SIAM Journal on Numerical Analysis 52~(1)
  (2014) 156--170.

\bibitem{blonigan_airfoil}
P.~J. Blonigan, Q.~Wang, E.~J. Nielsen, B.~Diskin, Least-squares shadowing
  sensitivity analysis of chaotic flow around a two-dimensional airfoil, AIAA
  Journal 56~(2) (2018) 658--672.

\bibitem{Pilyugin_1999}
S.~Y. Pilyugin, Shadowing in Dynamical Systems, Lecture Notes in Mathematics,
  Springer-Verlag, 1999.

\bibitem{Bowen1975-LimitDiffeomorphisms}
R.~Bowen, {{$\omega$}-Limit sets for Axiom A diffeomorphisms}, Journal of
  Differential Equations 18~(2) (1975) 333--339.

\bibitem{Hammel_et_al_1987}
S.~M. Hammel, J.~A. Yorke, C.~Grebogi, Do numerical orbits of chaotic dynamical
  processes represent true orbits?, Journal of Complexity 3~(2) (1987)
  136--145.

\bibitem{Sauer_Yorke_1991}
T.~Sauer, J.~A. Yorke, Rigorous verification of trajectories for the computer
  simulation of dynamical systems, Nonlinearity 4~(3) (1991) 961--979.

\bibitem{Sauer_et_al_1997}
T.~Sauer, C.~Grebogi, J.~A. Yorke, How long do numerical chaotic solutions
  remain valid?, Phys. Rev. Lett. 79 (1997) 59--62.

\bibitem{CHANDRAMOORTHY2021110389}
N.~Chandramoorthy, Q.~Wang, On the probability of finding nonphysical solutions
  through shadowing, Journal of Computational Physics 440 (2021) 110389.

\bibitem{bloniganMSS}
P.~J. Blonigan, Q.~Wang, Multiple shooting shadowing for sensitivity analysis
  of chaotic dynamical systems, Journal of Computational Physics 354 (2018)
  447--475.

\bibitem{Shawki_Papadakis_2019}
K.~Shawki, G.~Papadakis, A preconditioned multiple shooting shadowing algorithm
  for the sensitivity analysis of chaotic systems, J. Comput. Phys. 398 (2019)
  108861.

\bibitem{Kantarakias_Shawki_Papadakis_2020}
K.~Kantarakias, K.~Shawki, G.~Papadakis, Uncertainty quantification of
  sensitivities of time-average quantities in chaotic systems, Physical Review
  E 101 (2020) 022223.

\bibitem{NI201756}
A.~Ni, Q.~Wang, Sensitivity analysis on chaotic dynamical systems by
  non-intrusive least squares shadowing (nilss), Journal of Computational
  Physics 347 (2017) 56 -- 77.

\bibitem{NI2019690}
A.~Ni, C.~Talnikar, {Adjoint sensitivity analysis on chaotic dynamical systems
  by Non-Intrusive Least Squares Adjoint Shadowing (NILSAS)}, Journal of
  Computational Physics 395 (2019) 690 -- 709.

\bibitem{BLONIGAN2017803}
P.~J. Blonigan, Adjoint sensitivity analysis of chaotic dynamical systems with
  non-intrusive least squares shadowing, Journal of Computational Physics 348
  (2017) 803--826.

\bibitem{ni_2019}
A.~Ni, Hyperbolicity, shadowing directions and sensitivity analysis of a
  turbulent three-dimensional flow, Journal of Fluid Mechanics 863 (2019)
  644–669.

\bibitem{keefe_moin_kim_1992}
L.~Keefe, P.~Moin, J.~Kim, {The dimension of attractors underlying periodic
  turbulent Poiseuille flow}, Journal of Fluid Mechanics 242 (1992) 1–29.

\bibitem{vastano_moser_1991}
J.~A. Vastano, R.~D. Moser, Short-time {Lyapunov} exponent analysis and the
  transition to chaos in {Taylor–Couette} flow, Journal of Fluid Mechanics
  233 (1991) 83–118.

\bibitem{Hassanaly_Raman_2019}
M.~Hassanaly, V.~Raman, Lyapunov spectrum of forced homogeneous isotropic
  turbulent flows, Phys. Rev. Fluids 4 (2019) 114608.

\bibitem{Crisanti_et_al_1993}
A.~Crisanti, M.~H. Jensen, A.~Vulpiani, G.~Paladin, Intermittency and
  predictability in turbulence, Phys. Rev. Lett. 70 (1993) 166--169.

\bibitem{Mohan_et_al_2017}
P.~Mohan, N.~Fitzsimmons, R.~D. Moser, Scaling of {Lyapunov} exponents in
  homogeneous isotropic turbulence, Phys. Rev. Fluids 2 (2017) 114606.

\bibitem{pope_2000}
S.~B. Pope, Turbulent Flows, Cambridge University Press, 2000.

\bibitem{Pruett_2008}
C.~Pruett, Temporal large-eddy simulation: theory and implementation,
  Theoretical and Computational Fluid Dynamics 22 (2008) 275–304.

\bibitem{mckeon_sharma_2010}
B.~J. McKeon, A.~S. Sharma, A critical-layer framework for turbulent pipe flow,
  Journal of Fluid Mechanics 658 (2010) 336–382.

\bibitem{rigas_sipp_colonius_2021}
G.~Rigas, D.~Sipp, T.~Colonius, Nonlinear input/output analysis: application to
  boundary layer transition, Journal of Fluid Mechanics 911 (2021) A15.

\bibitem{padovan_otto_rowley_2020}
A.~Padovan, S.~E. Otto, C.~W. Rowley, Analysis of amplification mechanisms and
  cross-frequency interactions in nonlinear flows via the harmonic resolvent,
  Journal of Fluid Mechanics 900 (2020) A14.

\bibitem{moarref_jovanovic_2012}
R.~Moarref, M.~R. Jovanovic, {Model-based design of transverse wall
  oscillations for turbulent drag reduction}, Journal of Fluid Mechanics 707
  (2012) 205–240.

\bibitem{WANG20131}
Q.~Wang, Forward and adjoint sensitivity computation of chaotic dynamical
  systems, Journal of Computational Physics 235 (2013) 1 -- 13.

\bibitem{Pilyugin1999ShadowingSystems}
S.~Y. Pilyugin, {Shadowing in Dynamical Systems}, Vol. 1706 of Lecture Notes in
  Mathematics, Springer Berlin Heidelberg, 1999.

\bibitem{Lazarus_Thomas_2010}
A.~Lazarus, O.~Thomas, A harmonic-based method for computing the stability of
  periodic solutions of dynamical systems, Comptes Rendus Mécanique 338~(9)
  (2010) 510--517.

\bibitem{WereleyThesis}
N.~M. Wereley, {Analysis and control of linear periodically time varying
  systems}, Ph.D. thesis, Massachusetts Institute of Technology, Dept. of
  Aeronautics and Astronautics (1991).

\bibitem{LASAGNA2019119}
D.~Lasagna, A.~Sharma, J.~Meyers, Periodic shadowing sensitivity analysis of
  chaotic systems, Journal of Computational Physics 391 (2019) 119 -- 141.

\bibitem{HYMAN1986113}
J.~M. Hyman, B.~Nicolaenko, The \uppercase{K}uramoto-\uppercase{S}ivashinsky
  equation: A bridge between \uppercase{PDE}'s and dynamical systems, Physica
  D: Nonlinear Phenomena 18~(1) (1986) 113 -- 126.

\bibitem{BLONIGAN201416}
P.~J. Blonigan, Q.~Wang, Least squares shadowing sensitivity analysis of a
  modified {Kuramoto–Sivashinsky} equation, Chaos, Solitons \& Fractals 64
  (2014) 16 -- 25.

\bibitem{MUMPS}
P.~R. Amestoy, I.~S. Duff, J.-Y. L'Excellent, J.~Koster, A fully asynchronous
  multifrontal solver using distributed dynamic scheduling, SIAM Journal on
  Matrix Analysis and Applications 23~(1) (2001) 15--41.

\end{thebibliography}

\end{document}